\documentclass[11pt,a4paper]{article}
\usepackage{jheppub}
\usepackage{graphicx}
\pdfoutput=1
\usepackage{url}
\usepackage{cancel}
\usepackage{caption}
\usepackage{feynman}
\usepackage{bm}
\usepackage{hyperref}
\usepackage{enumerate}
\usepackage{amsmath}
\usepackage{amssymb}
\usepackage{listings}
\usepackage[toc, page]{appendix}
\usepackage{color}
\usepackage[section]{placeins}
\usepackage[percent]{overpic}

\def\d{ {D}}
\def\as{\alpha_s}
\def\nf{n_f}

\def\G0{{\bf \gamma^{(0)}_N}}
\def\G1{{\bf \gamma^{(1)}_N}}

\title{A fragmentation-based study of heavy quark production}
\author[a]{Giovanni Ridolfi,}
\author[b]{Maria Ubiali,}
\author[c,d]{Marco Zaro}
\affiliation[a]{Dipartimento di Fisica, Universit\`a di Genova \& INFN, Sezione di Genova
Via Dodecaneso 33, 16146 Genova, Italy}
\affiliation[b]{DAMTP, University of Cambridge, Wilberforce Road, Cambridge, CB3 0WA, United Kingdom}
\affiliation[c]{INFN, Sezione di Milano, Via Celoria 16, 20133 Milano, Italy}
\affiliation[d]{Nikhef, Theory Group, Science Park 105, 1098 XG,
  Amsterdam, The Netherlands}
\emailAdd{giovanni.ridolfi@ge.infn.it}
\emailAdd{M.Ubiali@damtp.cam.ac.uk}
\emailAdd{marco.zaro@mi.infn.it}

\abstract{Processes involving heavy quarks are a crucial
 component of the LHC physics program, both by themselves and as
 backgrounds for Higgs physics and new physics searches. 
 In this work, we critically reconsider the validity of the widely-adopted approximation 
 in which heavy quarks are generated at the matrix-element level, with
 special emphasis on the impact of the collinear logarithms associated
 with final-state heavy quark and gluon splittings. 
Our study, based on a perturbative fragmentation-function approach,
explicitly shows that neglecting the resummation of collinear logarithms may yield 
inaccurate predictions, in particular when observables exclusive in
the heavy quark degrees of freedom are considered. Our findings motivate the 
use of schemes which encompass the resummation of final-state collinear logarithms.}
\arxivnumber{1911.01975}
\subheader{\small DAMTP-2019-27\\NIKHEF/2019-044\\TIF-UNIMI-2019-20}

\begin{document}

\maketitle
\flushbottom

\section{Introduction}
The production of heavy quarks in association with other particles 
at hadron colliders represents a crucial testing ground for our understanding 
of perturbative Quantum Chromodynamics (QCD) in the presence of several energy
scales. This class of processes is governed by at least two scales, 
namely the heavy-quark mass $m$ and the (invariant) 
mass $M$ of the particle(s) produced along with the heavy quark. 
In these cases, large collinear logarithms of the ratio $\frac{M}{m}$ may jeopardise 
the convergence of the perturbative expansion of relevant theoretical 
predictions. 
Fortunately, the impact of these logarithmic contributions can be controlled
by resumming them to all orders in $\as$, via a scheme 
in which the heavy quark mass $m$ is neglected at the level of the matrix element. 
Such a scheme is 
often referred to as massless or five-flavour scheme (5FS), in case the heavy quark
is identified with the bottom quark.
As far as heavy quarks in the initial state are concerned, this procedure amounts to
introduce a suitable parton distribution for the heavy quark. An analogous procedure for 
heavy quarks in the final state involves the use of fragmentation functions, and is the subject of the present work. A scheme in which 
the heavy quark is produced at the matrix-element level and is not treated on the
same footings as the light quarks is dubbed as massive scheme or four-flavour scheme (4FS).
The resummation of powers of $\log(M/m)$ in a 5FS is performed by solving the 
evolution equations
(usually referred to as Dokshitzer-Gribov-Lipatov-Altarelli-Parisi, 
or DGLAP, equations),
at the price of discarding power corrections of ${\cal O}(m^2/M^2)$, and thus of
yielding less accurate theoretical predictions for the observables related 
to the heavy-quark degrees of freedom.

In~\cite{Maltoni:2012pa, Lim:2016wjo}, it has been shown that, 
for processes in which the heavy quarks 
(more specifically bottom quarks) are dominantly produced via 
initial-state (spacelike) splittings, the theoretical predictions in
4FS are not spoiled by initial-state collinear logarithms. This is due to two main factors, one of
dynamical and the other of kinematical origin. The first is that 
the effects of the resummation of the initial-state collinear
logarithms are relevant mainly at large $x$ and, in general,
keeping only the explicit logs appearing at NLO in the 4FS is a good 
approximation. The second reason is that the scale which appears in the 
collinear logarithms turns out to be proportional to the hard scale of
the process but is suppressed by universal phase space factors that, 
at hadron colliders, reduce the size of the
logarithms for processes taking place. 
This result makes it not only possible, but also 
advisable -- owing to the better perturbative description of 
the differential observables involving the heavy quark(s) --
to employ the 4FS for the exclusive description of these processes. 
This has been shown explicitly to be the case in single-top 
production~\cite{Campbell:2009ss,Frederix:2012dh}, 
$b\bar b H$~\cite{Maltoni:2003pn,Boos:2003yi,Harlander:2003ai,Dawson:2003kb,Dawson:2004sh,Dawson:2005vi,Harlander:2011aa,Harlander:2014hya,Wiesemann:2014ioa,Hartanto:2015uka,Jager:2015hka,Krauss:2016orf} and 
$b\bar b Z/\gamma$ production~\cite{FebresCordero:2008ci,Cordero:2009kv,Frederix:2011qg,Hartanto:2013aha,Cordero:2015sba,Krauss:2016orf,Figueroa:2018chn,Bagnaschi:2018dnh}, 
and also for processes predicted by extensions of the Standard Model (SM), 
such as heavy charged Higgs boson production in a two-Higgs 
doublet model or in 
supersymmetry~\cite{Zhu:2001nt,Gao:2002is,Plehn:2002vy,Berger:2003sm,Peng:2006wv,Dittmaier:2009np,Weydert:2009vr,Klasen:2012wq,Flechl:2014wfa,Degrande:2015vpa,Degrande:2016hyf}. 
On the other hand, the calculations of the total rates in the 5FS
display a faster perturbative convergence and exhibit a smaller scale uncertainty 
associated with missing higher orders. 
Methods that combine the 4F and
the 5F schemes, retaining the advantages of both, are
actually available, but they are generally
tailored to a few specific observables. 
The FONLL scheme, first proposed for the transverse momentum spectrum
of bottom quarks produced in hadronic
collisions~\cite{Cacciari:1998it}, has the advantage of being
universally applicable and of allowing one to combine
4FS and 5FS calculations performed at any perturbative
order. The formulation of the FONLL scheme has been extended to 
deep-inelastic scattering (DIS)~\cite{Forte:2010ta} and adapted 
to the computation of the total cross section for Higgs and $Z$ production in
bottom-quark fusion~\cite{Forte:2015hba,Forte:2016sja,Forte:2018ovl}.
Various recent attempts to consistently include both the resummation
of initial-state collinear logarithms and mass effects also for differential and
parton-shower matched observables have been recently put forward, see for example 
the five-flavour-massive scheme proposed in~\cite{Krauss:2016orf,Krauss:2017wmx} 
or a similar approach based on multi-jet merging~\cite{Hoche:2019ncc}. 
Improvements at the inclusive and at the differential level are on-going.
Finally, consistent b-quark PDFs to be used in association with 
massive initial states have also been defined~\cite{Forte:2019hjc}, thus 
allowing the bottom quark to be endowed with a standard PDF satisfying DGLAP  
evolution equations, yet treating it as massive in hard matrix elements.

While initial-state collinear logarithms have been studied in details in the above-mentioned 
literature, the situation is much less clear for processes in which final-state
(timelike) splittings into heavy quarks contribute significantly to the process. It has to be mentioned
that, for what concerns the production of flavoured jets, higher-order corrections are generally not finite (or they are logarithmically enhanced 
in a massive scheme)
unless dedicated jet algorithms are employed, see Ref.~\cite{Banfi:2006hf}. Processes featuring final-state splittings into heavy quarks include
$b\bar b W$ production~\cite{FebresCordero:2006nvf,Campbell:2008hh,Cordero:2009kv,Badger:2010mg,Oleari:2011ey,Frederix:2011qg,Caola:2011pz,Reina:2011mb}, 
the top-mediated contribution 
to $b\bar b H$ production~\cite{Deutschmann:2018avk}, 
$t\bar t b\bar b$ production~\cite{Bevilacqua:2009zn,Bredenstein:2010rs,Cascioli:2013era,Garzelli:2014aba,Bevilacqua:2017cru,Jezo:2018yaf} and multi-$b$ 
final states~\cite{Greiner:2011mp,Bevilacqua:2013taa} 
(mostly relevant for di- or triple-Higgs 
searches~\cite{deLima:2014dta,Alves:2019igs,Papaefstathiou:2019ofh}). 
While the importance of the resummation of collinear logarithms has
been partially investigated for $Q\to Qg$
splittings~\cite{Mele:1990cw,Oleari:1997az,Cacciari:1998it}, 
no assessment of the impact of the 
logarithms of $\frac{M}{m}$ exists to date, as far as the $g\to Q\bar
Q$ splittings are concerned.

An interesting process which involves bottom quarks in the final state is the production
of  $t\bar t b\bar b$. This process is an important background to Higgs and top
associated production, a unique probe of the Yukawa coupling between the Higgs scalar and top 
quarks, and therefore it is of great relevance for present-day
analyses~\cite{Aaboud:2017rss,Sirunyan:2018mvw}.
Different tools are available to simulate this process in a 4FS, including
NLO QCD corrections and matching with parton showers. However,
even when tuned comparisons are performed~\cite{deFlorian:2016spz},
the predictions obtained by different tools display rather large differences,
which have a dominant impact on the systematic uncertainty in the determination of the top
Yukawa coupling\footnote{Ongoing work and preliminary results are 
    available here: \url{https://twiki.cern.ch/twiki/bin/view/LHCPhysics/ProposalWwbbbb}.}. 
The work necessary to improve this situation by increasing the
perturbative order of the computation is not straightforward. 
Given the high multiplicity and the number of scales involved in this kind of processes, 
the NNLO corrections in the 4FS are very hard to compute.
On the other hand, NLO QCD predictions for $t\bar t b\bar b$ plus one light jet have 
recently become available~\cite{Buccioni:2019plc}. Assuming that the
distortion due to parton showers is small, these calculations
could help to validate the light-jet spectrum. 
If the resummation of collinear logarithms associated with the
final-state splittings of gluons and bottom quarks is found to have a
strong impact on this observable, then a matched calculation could
solve the observed discrepancies. It is the purpose of the present work
to make a first step in this direction.

In this paper, we 
assess the impact of missing powers of  $\log\frac{M}{m}$ associated
to final-state splittings
by means of fragmentation functions (FFs). Heavy quark FFs can be computed in perturbation 
theory in QCD, starting from initial conditions at a reference scale 
$\mu_0$ and employing 
the timelike DGLAP evolution equations to evolve them up to any other scale. 
Initial conditions for the gluon- and heavy-quark-initiated fragmentation into a heavy quark 
are known at order $\as$~\cite{Mele:1990yq,Mele:1990cw} and have been computed  
at order $\as^2$~\cite{Melnikov:2004bm,Mitov:2004du}, while the DGLAP 
evolution equation is implemented
in public codes such as {\tt QCDnum}~\cite{Botje:2010ay}, 
{\tt ffevol}~\cite{Hirai:2011si},  
{\tt APFEL}~\cite{Bertone:2013vaa} 
or {\tt MELA}~\cite{Bertone:2015cwa}, up to NNLL 
logarithmic accuracy. The codes have been benchmarked in~\cite{Bertone:2015cwa}. 
An approach based on FFs will enable us to study the dynamics
of the bottom fragmentation in details 
and in an isolated environment. In particular, the importance
of the resummation of potentially large logarithmic contributions can be assessed by comparing
resummed predictions to the truncation of the FF at a given order in $\as$,
and the impact of the resummation of sub-leading logarithms can be studied
up to NNLL accuracy. 
It must be stressed that, while the importance of resumming collinear logarithms in bottom-quark initiated fragmentation has been known
for a long time at NLL~\cite{Mele:1990cw,Oleari:1997az,Cacciari:1998it}, 
much less attention has been devoted to the role of
gluon-initiated fragmentation to heavy quarks (one exception is
Ref.~\cite{Helenius:2018uul} in which the case of charm meson production
was studied). Such a negligence was justified
 the past by the sub-dominant importance of this mechanism at LEP and at Tevatron, 
but this is no longer the case at the LHC for the $g\to b \bar b$ splitting,
and will not be the case for the $g\to t \bar t$ splitting at future colliders.

The paper is organised as follows. We review the details of timelike DGLAP evolution
in Sec.~\ref{sec:dglap}, where we also discuss how to truncate the evolution
at a given order in $\as$. In Sec.~\ref{sec:setup}, we discuss 
the setup employed for this computation, while results are
presented in Sec.~\ref{sec:results}. In the light of our results,
in Sec.~\ref{sec:simul} we comment on how to simulate processes in which $b$ quarks 
are dominantly created in final-state splittings. In Sec.~\ref{sec:hqm} we link our 
findings to those obtained in the context of heavy quark multiplicity estimates. 
We draw our conclusions in Sec.~\ref{sec:conc},
where we also discuss future outlooks of our work. In Appendix~\ref{app:truncate}, we provide 
supplementary material, namely the explicit expressions of the truncated 
FFs up to order $\as^3$ together
with the discussion of their numerical validation.

\section{Timelike DGLAP evolution}
\label{sec:dglap}
In this Section we review the formalism of scale evolution for fragmentation functions,
with the main purpose of fixing notations and conventions.
We also set the ground for the derivation of explicit formulae for heavy-quark fragmentation functions at fixed order
up to order $\as^3$, which are reported in Appendix~\ref{app:truncate}. 

\subsection{Strong coupling constant}
We adopt the following notation for the evolution of the running coupling
constant $\as(\mu)$ of strong interactions:
\begin{equation}
\frac{d\as(\mu)}{dt}=\beta(\as(\mu)) \qquad \beta(\as)=-\as^2\left(b_0+b_1\as\right)+{\cal
O}(\as^4),
\label{eq:alpharge}
\end{equation}
where $t=\log\frac{\mu^2}{\mu_0^2}$, $\mu_0$ is a fixed reference scale, and
\begin{eqnarray}
&&b_0=\frac{11C_A-4\nf T_F}{12\pi}\qquad b_1=\frac{17C_A^2-10C_A\nf T_F-6C_F\nf T_F}{24\pi^2}.
\end{eqnarray}
As usual, $T_F=\frac{1}{2}$, $C_A=3$ and $C_F=\frac{4}{3}$ for three-colour QCD.
The number of active flavours $n_f$ will always be set to 5.

We will need  the expansion of $\alpha_s(\mu)$ in powers of $\alpha_s(\mu_0)$ 
truncated at ${\cal O}(\as^3)$, which is given by
\begin{equation}
 \as(\mu)=\as(\mu_0)-\as^2(\mu_0)b_0t+\as^3(\mu_0)(b_0^2t^2-b_1t).
 \label{eq:asexp}
\end{equation}
The truncated expansion of $\alpha_s(\mu_0)$ in terms of $\alpha_s(\mu)$ can be trivially obtained by swapping $\mu$ and
$\mu_0$ in the above equation obtaining
\begin{equation}
 \as(\mu_0)=\as(\mu)+\as^2(\mu)b_0t+\as^3(\mu)(b_0^2t^2 + b_1t).
 \label{eq:as0exp}
\end{equation}

\subsection{Fragmentation functions}
We consider the differential cross section $\frac{d\sigma}{dx}$ for a generic process with a heavy quark $Q$ of mass $m$ in the final state, where $x$ is the energy fraction
carried by the heavy quark:
\begin{equation}
x=\frac{E_Q}{E},
\end{equation}
where $E_Q$ is the energy of the heavy quark, and $E$ the energy of the parton originating from the matrix element. Then,
standard factorisation implies
\begin{equation}
\frac{d\sigma}{dx}(x,E,m)=\sum_i\int_x^1\,\frac{dz}{z}\,
\frac{d\hat\sigma_i}{dz}(z,E,m)\,D_i\left(\frac{x}{z},\mu,m\right),
\end{equation}
where $\frac{d\hat\sigma_i}{dz}$ is the partonic cross section
for parton $i$ in the final state with energy fraction $z$, and $D_i(x,\mu,m)$ is the fragmentation function of parton $i$ into the heavy quark.

The fragmentation functions depend on the factorisation scale $\mu$
according to the evolution equations
\begin{equation}
\mu^2\frac{dD_i}{d\mu^2}(x,\mu,m)=\sum_j \,
\int_x^1\,\frac{dz}{z}\,P_{ij}\left(\frac{x}{z},\as(\mu)\right)
\,D_j(z,\mu,m).
\end{equation}
The timelike splitting functions $P_{ij}$ have a power expansion in $\as$, whose
coefficients have been computed up to NNLO, and can be found 
in~\cite{Mitov:2006ic,Moch:2007tx,Almasy:2011eq}.
Note that, in the case of timelike evolution, we have
\begin{equation}
P_{12}(x,\as)=P_{gq}(x,\as);\qquad
P_{21}(x,\as)=P_{qg}(x,\as),
\end{equation}
contrary to what happens in the spacelike case. The timelike splitting functions are the same
as the spacelike ones at LO, while they differ at NLO and
higher. 

The DGLAP evolution equations are conveniently solved for Mellin-transformed quantities, because Mellin transformation turns the integro-differential DGLAP equations into ordinary differential equations. We define the Mellin transform $f(N)$ of a generic function $f(x)$
by
\begin{equation}
  f(N)=\int_0^1dx\,x^{N-1}f(x).
\end{equation}
We will use the same symbol for a function and its Mellin transform; this does not lead to
confusion, as long as functional arguments are explicitly indicated.
We rewrite the timelike DGLAP evolution equations as
\begin{equation}
\label{eq:tDGLAP}
\frac{d\d_i}{dt}(N,t,m)\,=\,\sum_j\,\gamma_{ij}(N,t)\d_j(N,t,m),
\end{equation}
where
\begin{eqnarray}
\gamma_{ij}(N,t)
&=&\int_0^1dz\,z^{N-1}P_{ij}(z,\as(\mu))
\notag\\
&=&\frac{\as(\mu)}{4\pi}\gamma_{ij}^{(0)}(N)+\left(\frac{\as(\mu)}{4\pi}\right)^2\gamma_{ij}^{(1)}(N)  +
\left(\frac{\as(\mu)}{4\pi}\right)^3\gamma_{ij}^{(2)}(N)+ {\cal O}(\as^4).
\label{eq:gammaexp}
\end{eqnarray}
The LO singlet timelike anomalous dimensions $\gamma^{(0)}_{ij}$ 
are given by
\begin{align}
\gamma_{qq}^{(0)}(N)&=C_F\left[3-4S_1(N)+\frac{2}{N(N+1)}\right]
\label{eq:gammaqq}
\\
\gamma_{gq}^{(0)}(N)&=2\nf C_F\frac{2(N^2+N+2)}{(N-1)N(N+1)}
\label{eq:gammagq}
\\
\gamma_{qg}^{(0)}(N)&=T_F\frac{2(N^2+N+2)}{N(N+1)(N+2)}
\label{eq:gammaqg}
\\
\gamma_{gg}^{(0)}(N)&=C_A\left[\frac{11}{3}-4S_1(N)+\frac{4}{N(N-1)}
+\frac{4}{(N+1)(N+2)}\right]
-\frac{4}{3}T_F\nf,
\label{eq:gammagg}
\end{align}
where $S_1(N)$ is the harmonic sum of order 1 as defined in~\cite{Vermaseren:1998uu}.
The anomalous dimensions $\gamma^{(1)}_{ij}$ 
were calculated in~\cite{Floratos:1981hs,Gluck:1992zx}\footnote{Note that an error in the computation of the splitting functions in~\cite{Floratos:1981hs} 
was fixed in~\cite{Gluck:1992zx}, due to the choice of an unphysical factorisation scheme.} and $\gamma^{(2)}_{ij}$  in~\cite{Moch:2007tx,Almasy:2011eq}. 

\subsection{Initial conditions}
We will need suitable initial conditions for the fragmentation functions.
The perturbative initial conditions have been computed
at order $\as$ in Refs.~\cite{Mele:1990yq,Oleari:1997az} and $\as^2$ in Refs.~\cite{Mitov:2004du,Melnikov:2004bm}:
\begin{align}
\d_b(N,0,m) &= 1 +  \frac{\as(\mu_0)}{4\pi} d^{(1)}_b(N,\mu_0,m)  + \left(\frac{\as(\mu_0)}{4\pi}\right)^2 d^{(2)}_b(N,\mu_0,m) \,,  \nonumber\\
\d_{\bar{b}}(N,0,m) &= \left(\frac{\as(\mu_0)}{4\pi}\right)^2 d^{(2)}_{\bar{b}}(N,\mu_0,m) \,,  \nonumber\\
\d_g(N,0,m) &=  \frac{\as(\mu_0)}{4\pi}d^{(1)}_g(N,\mu_0,m)  + \left(\frac{\as(\mu_0)}{4\pi}\right)^2 d^{(2)}_g(N,\mu_0,m) \,, \nonumber \\
\d_q(N,0,m) &= \d_{\bar{q}}(N,0,m) = \left(\frac{\as(\mu_0)}{4\pi}\right)^2 d^{(2)}_q(N,\mu_0,m) \,. \label{eq:init}
\end{align}
Initial conditions at order $\as^3$ are currently unknown, and will be neglected in the following.

It is interesting to notice that the initial condition for the $b$ quark fragmentation function contains a non-logarithmic term already at NLO, namely
\begin{align}
d_{b}^{(1)}(N,\mu_0,m)&=2C_F\Bigg[\left(\frac{3}{2}+\frac{1}{N(N+1)}-2S_1(N)\right)\log\frac{\mu_0^2}{m^2}-2S^2_1(N)\notag\\
&\qquad+\frac{2S_1(N)}{N(N+1)}-\frac{2}{(N+1)^2}-2S_2(N)+2-\frac{1}{N(N+1)}+2S_1(N)\Bigg],
\label{eq:initb}
\end{align}
contrary to the case of the $b$ quark PDF in the
spacelike evolution~\cite{Buza:1995ie,Buza:1996wv}.
The initial-scale gluon fragmentation function has instead only a logarithmic term that vanishes when $\mu_0=m$:
\begin{align}
d_{g}^{(1)}(N,\mu_0,m)&=\gamma_{qg}^{(0)}\log\frac{\mu_0^2}{m^2}.
\label{eq:initg}
\end{align}

It is customary to separate singlet from non-singlet evolution in the DGLAP equations.
To this purpose, we define the combinations
\begin{align}
\label{eq:initcond}
\d_\Sigma&=\sum_{i=1}^{\nf }\,\d_{q_i}^+\notag \\
\d_{V_i} &= \d_{q_i}-\d_{\bar{q}_i}\qquad\qquad\qquad\qquad\qquad\qquad\qquad\qquad i=1,...,\nf  \notag \\
\d_{T_3}&=(\d_u+\d_{\bar{u}})-(\d_d+\d_{\bar{d}})\notag \\
\d_{T_8}&=(\d_u+\d_{\bar{u}})+(\d_d+\d_{\bar{d}})-2(\d_s+\d_{\bar{s}}) \notag \\
\d_{T_{15}}&=(\d_u+\d_{\bar{u}})+(\d_d+\d_{\bar{d}})+(\d_s+\d_{\bar{s}})-3(\d_c+\d_{\bar{c}}) \\
\d_{T_{24}}&=(\d_u+\d_{\bar{u}})+(\d_d+\d_{\bar{d}})+(\d_s+\d_{\bar{s}})+(\d_c+\d_{\bar{c}})-4(\d_b+\d_{\bar{b}})\notag
\end{align}
with the valence contributions evolving according to the non-singlet
($V$) timelike evolution equations and the triplet contributions evolving 
according to the non-singlet
($+$) timelike evolution equations. The evolution of the
singlet combination $\d_\Sigma$ is coupled with the gluon. 
The bottom quark fragmentation is given by
\begin{equation}
\d_b=\frac{\d_{\Sigma}-\d_{T_{24}}+\nf \d_{V_b}}{2\nf },
\label{eq:combination}
\end{equation}
and the non-vanishing initial conditions are given by
\begin{align}
\d_{\Sigma}(N,0,m)&= 1 + \frac{\as(\mu_0)}{4\pi} d^{(1)}_b(N,\mu_0,m) \\
& \qquad + \left(\frac{\as(\mu_0)}{4\pi}\right)^2 (d^{(2)}_b(N,\mu_0,m) +  d^{(2)}_{\bar{b}}(N,\mu_0,m)\notag\\
& \qquad \qquad \qquad \qquad +2(\nf -1)\,d^{(2)}_q(N,\mu_0,m))\notag \\
\d_{V_b}(N,0,m) &= \d_b-\d_{\bar{b}} = 1 + \frac{\as(\mu_0)}{4\pi}d^{(1)}_b(N,\mu_0,m)\\
&\qquad\qquad\qquad\quad+\left(\frac{\as(\mu_0)}{4\pi}\right)^2 (d^{(2)}_b(N,\mu_0,m) -  d^{(2)}_{\bar{b}}(N,\mu_0,m))\notag \\
\d_{T_{24}}(N,0,m)&= 2(\nf -1)\left(\frac{\as(\mu_0)}{4\pi}\right)^2 d^{(2)}_q(N,\mu_0,m)\\
&- (\nf -1)\Bigg[1 + \frac{\as(\mu_0)}{4\pi}d^{(1)}_b(N,\mu_0,m) \notag\\
& \qquad\qquad + \left(\frac{\as(\mu_0)}{4\pi}\right)^2 \left(d^{(2)}_b(N,\mu_0,m) + d^{(2)}_{\bar{b}}(N,\mu_0,m)\right)\Bigg].\notag
\end{align}

To determine $\d_b$ and $\d_g$ up to NNLO and their expansions up to ${\cal O}(\as^3)$,
we need solutions of the evolution equations for both the singlet and the
triplet combinations
\begin{align}
\label{eq:dglapNS}
\frac{d\d_T(N,t,m)}{dt}\,&=\,\gamma_{+}(t)\d_T(N,t,m)\\
\frac{d\d_V(N,t,m)}{dt}\,&=\,\gamma_{V}(t)\d_V(N,t,m),
\end{align}
for $\d_{T_{24}}$ and $\d_{V_b}$ respectively, and
\begin{align}
\label{eq:dglapSG}
\frac{d}{dt}\begin{pmatrix}
\d_{\Sigma}\\
\d_{g}
\end{pmatrix}&=\,
\begin{pmatrix}
\gamma_{qq} &   \gamma_{gq}\\
\gamma_{qg} &   \gamma_{gg}
\end{pmatrix}
\begin{pmatrix}
\d_{\Sigma}\\
\d_{g}
\end{pmatrix}.
\end{align}
Note that Eq.~\eqref{eq:dglapSG} differs from Eq.~(2.4) of Ref.~\cite{Bertone:2015cwa} only formally, 
because in the latter $\gamma_{qg}$ is the element of the spacelike singlet matrix evolution, that
features a factor $2\nf$.\footnote{In MELA, particularly in Eq.~(2.4), $P_{ij}$ do not refer to the usual splitting 
functions, rather they refer to the $ij$
entries of the singlet spacelike evolution matrix (Eq.~(4.100) of Ref.~\cite{Ellis:1991qj}). 
The factors $2\nf$ and $1/2\nf$ are explained in Appendix A2 of Ref.~\cite{Gluck:1992zx}.}

%
%

\subsection{Truncated solution of DGLAP equation in Mellin space}
The DGLAP equations are usually solved in order to resum large logarithmic contributions;
the fragmentation functions are evolved from
a reference scale $\mu_0$ to a generic scale $\mu$ through an evolution operator,
which in turn is given by an expansion in powers of $\as(\mu_0)$
with $\as(\mu_0)\log\frac{\mu^2}{\mu_0^2}$ fixed.
Here, we would like to compare such a logarithmic (resummed) expansion with a truncated one, that is, a
solution expressed as a power series in $\as(\mu)$, 
up to a certain order.

We rewrite Eq.~\eqref{eq:tDGLAP} in matrix form, and with some of the functional arguments omitted,
to keep notation simple:
\begin{equation}
\label{eq:tDGLAPgrm}
\frac{d\d(t)}{dt}=\gamma(t)\d(t).
\end{equation}
Eq.~(\ref{eq:tDGLAPgrm}) has the solution
\begin{equation}
\d(t)=U(t,0)\d(0),
\end{equation}
with 
\begin{eqnarray}
U(t,0)&=&I+\int_0^t dt_1\,\gamma(t_1)
+\int_0^t dt_1\,\gamma(t_1)\int_0^{t_1} dt_2\,\gamma(t_2)+\ldots
\notag\\
&=&\sum_{n=0}^\infty\frac{1}{n!}\int_0^t dt_1\ldots\int_0^t dt_n\,
T\left[\gamma(t_1)\ldots \gamma(t_n)\right],
\end{eqnarray}
and $T$ is the time-ordering operator.
The matrix $\gamma$ has a Taylor expansion in $\as$, given in Eq.~\eqref{eq:gammaexp}, which starts at order $\as$.
Therefore,  it is easy to truncate
the expansion of $U(t,0)$ to any given order. Given that we are interested
in the solution up to NNLO, we keep terms up to order $\as^3$ in $U(t,0)$. We find
\begin{eqnarray}
U_{\rm NNLO}(t,0)&=&
I+\gamma^{(0)}\int_0^t dt_1\,\frac{\as(\mu_1)}{4\pi}
\nonumber\\
&&+\frac{1}{2}
\left[\gamma^{(0)}\int_0^t dt_1\,\frac{\as(\mu_1)}{4\pi}\right]^2
+\gamma^{(1)}\int_0^tdt_1\,\left(\frac{\as(\mu_1)}{4\pi}\right)^2
\nonumber\\
&&+\gamma^{(0)}\gamma^{(1)}\int_0^tdt_1\,\frac{\as(\mu_1)}{4\pi}
\int_0^{t_1}dt_2\,\left(\frac{\as(\mu_2)}{4\pi}\right)^2
\nonumber\\
&&+\gamma^{(1)}\gamma^{(0)}\int_0^tdt_1\,\left(\frac{\as(\mu_1)}{4\pi}\right)^2
\int_0^{t_1}dt_2\,\frac{\as(\mu_2)}{4\pi}
\nonumber\\
&&+\frac{1}{6}\left[\gamma^{(0)}\int_0^tdt_1\,\frac{\as(\mu_1)}{4\pi}\right]^3+\gamma^{(2)}\int_0^tdt_1\,\left(\frac{\as(\mu_1)}{4\pi}\right)^3,
\label{eq:unnlo}
\end{eqnarray}
where $t_1=\log\frac{\mu_1^2}{\mu_0^2}$, $t_2=\log\frac{\mu_2^2}{\mu_0^2}$.
Next, $\as(\mu_1),\as(\mu_2)$ are expanded in powers of $\as(\mu_0)$ as in Eq.~\eqref{eq:asexp},
and the integrals easily performed. Finally, $\as(\mu_0)$ is re-expressed in terms of $\as(\mu)$, according to Eq.~\ref{eq:as0exp}.

In the following, we will call LO, NLO and NNLO truncated FFs, respectively the expression obtained by evolving the initial 
conditions of Eq.~\eqref{eq:init} with the evolution operator in Eq.~\eqref{eq:unnlo}, and retaining terms up to order $\as(\mu)$, 
$\as^2(\mu)$ and $\as^3(\mu)$ respectively\footnote{Note that $D_b$ starts at order $\as^0(\mu)$.}. The full expressions are reported in Appendix~\ref{app:truncate}.


\section{Setup of the computation}
\label{sec:setup}
The results presented in this paper are obtained by means of a private computer code which links the
public {\tt MELA} (Mellin Evolution LibrAry) library~\cite{Bertone:2015cwa}. 
\texttt{MELA} is an evolution program in Mellin space, developed specifically to provide a simple and user-friendly framework 
complementary to (and also serving as a validation of) the 
code \texttt{APFEL}~\cite{Bertone:2013vaa}, which works in $x$ space. 

For the running of $\alpha_s$, we use the routines
implemented in {\tt MELA} with $\alpha_s(M_Z)=0.11856$ and
$M_Z=91.187$ GeV, that solve the renormalisation group equation for
$\alpha_s(\mu)$
consistently with the DGLAP timelike equations. 
The charm and bottom thresholds are set to
$m_c=1.4142$ GeV and $m_b= 4.7$ GeV respectively. 
The top quark mass $m_t$ is set to infinity, so that $n_f=5$ at all
scales. 
The timelike splitting functions at LO, NLO and NNLO in the $N$ space are
taken directly from {\tt MELA}.  Note that, due to the complexity of the expressions entering the NNLO splitting functions, 
{\tt MELA} implements the
approximate representation of Ref.~\cite{vogt}. It was checked in~\cite{Almasy:2011eq} that, except for very small
values of $x$, such approximate expressions deviate from the exact ones by less than one part
in a thousand.
The $N$-space solution of the timelike DGLAP evolution equation at
LL, NLL and NNLL are also taken from  {\tt MELA}. 
{\tt MELA} implements the analytical solutions of 
the DGLAP evolution equations as in {\tt PEGASUS} ~\cite{Vogt:2004ns},
both the truncated solution and the iterated solution. In the former,  
the resummed solution to the DGLAP equations in $N$ space is exact up to
terms of higher orders in the perturbative expansion with respect to
the order of the DGLAP solution. In the iterated solution, all orders
are kept in the solution of the DGLAP equation. The N$^m$LL solutions 
differ in terms of order $n>m$. In our case, we have verified that the
effect of the resummation of collinear logarithms does not depend on
the settings of the solution of the DGLAP evolution equation.

The initial perturbative conditions have been implemented in our 
own code in the $N$ space up to order $\as^2$ by numerically 
Mellin-transforming the $x$-space expressions of
Refs.~\cite{Melnikov:2004bm,Mitov:2004du} for real $N$. Di- and tri-logarithms appearing
in these expressions are evaluated with {\tt Chaplin}~\cite{Buehler:2011ev}.
Since we are currently lacking an analytically-continued Mellin
transform of the ${\cal O}(\as^2)$ terms of the initial conditions\footnote{More specifically:
we succeeded to obtain analytically-continued Mellin transforms for 
$d_b^{(2)}$, $d_{\bar b}^{(2)}$, $d_q^{(2)}$, following
the results and algorithms of 
Refs.~\cite{Blumlein:1997vf,Blumlein:1998if,Blumlein:2000hw,Blumlein:2003gb,Blumlein:2009ta}, 
(as implemented in the codes {\tt ancont} and {\tt ancont1}),
while for $d_g^{(2)}$ some terms are still missing. We plan to report
on the complete analytic continuation of the initial conditions in a following publication.}
the evolved expressions cannot be inverted to $x$ space if 
these initial conditions are included.
As their impact in Mellin space is found to be mild and rather flat in
the $N$ space, as we will show explicitly in Sec.~4, we argue that
they will not play an important role in the $x$ space.
The numerical inversion of the $N$-space truncated and
resummed fragmentation functions from
$N$ space back to $x$ space is performed 
by means of an implementation of the Mellin inversion based on the 
Talbot-path algorithm~\cite{doi:10.1002/nme.995}. Matching conditions
are implemented in the treatment of flavour threshold crossing in the
evolution of the fragmentation functions.


\section{Impact of collinear resummation and results}
\label{sec:results}
In this Section, we present results for the resummed and truncated FFs at different mass scales, in
Mellin ($N$) space as well as in the physical space of the energy
fraction $x$ carried by the heavy quark.

By comparing the truncated and resummed predictions for the FFs for different values of the scales
$\mu$ and $\mu_0$ and at different orders, we can: 
\begin{enumerate}[i)]
    \item assess the typical size of the effects due to the resummation of final-state collinear logarithms, in particular
        with respect to an approximation in which only logarithms up to a given order in perturbation theory are included;
    \item compare the behaviour of the bottom-quark and gluon initiated FF;
    \item determine the importance of the inclusion of initial
      conditions at order $\as^2$, computed 
in~\cite{Melnikov:2004bm,Mitov:2004du}.
\end{enumerate}
In particular, the last point and the importance of the gluon FF have
been neglected  so far in the literature, see
e.g.~\cite{Oleari:1997az}. 

\begin{figure}[h!]
    \centering
    \includegraphics[width=\textwidth]{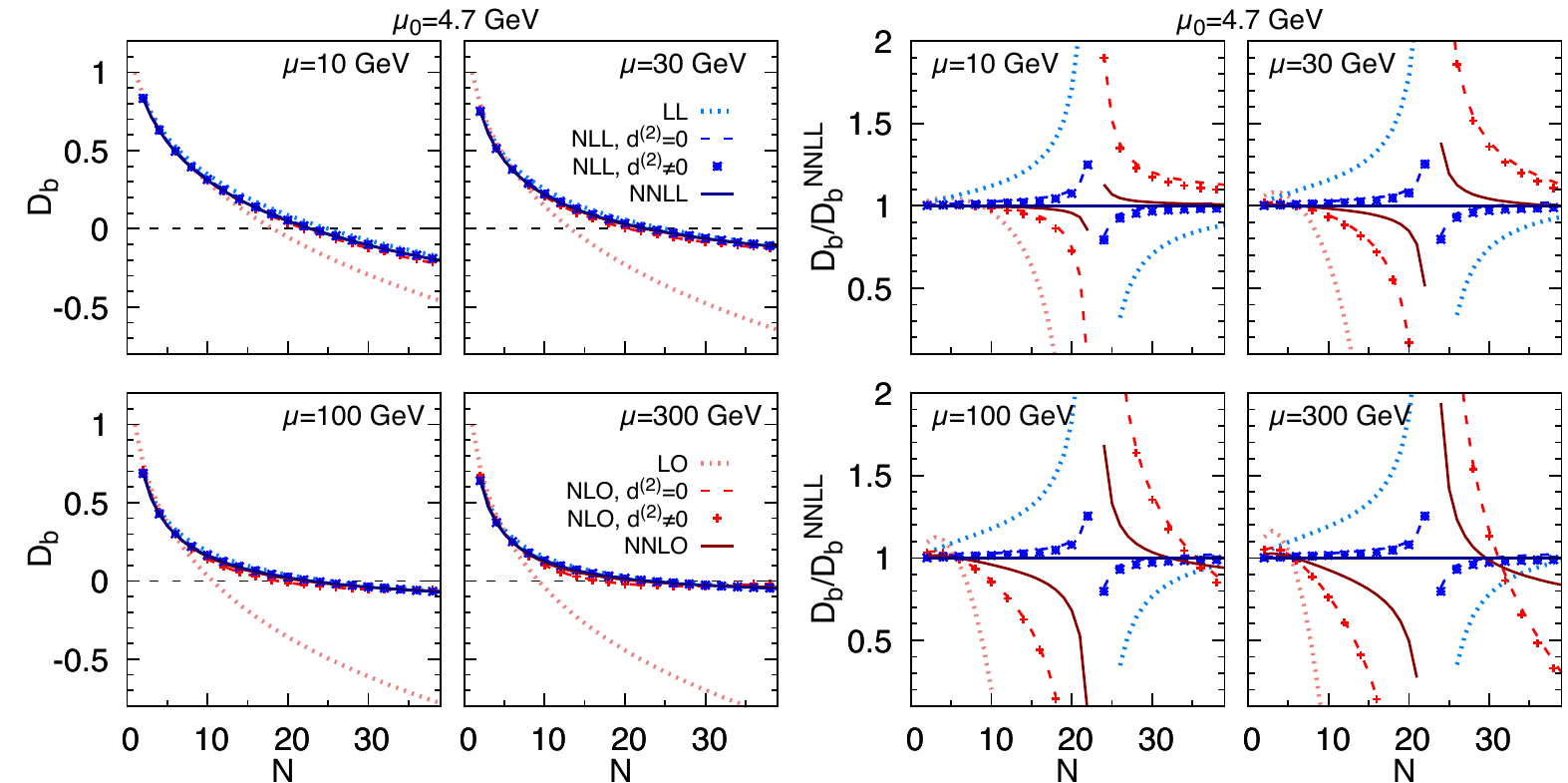}\\[5mm]
    \includegraphics[width=\textwidth]{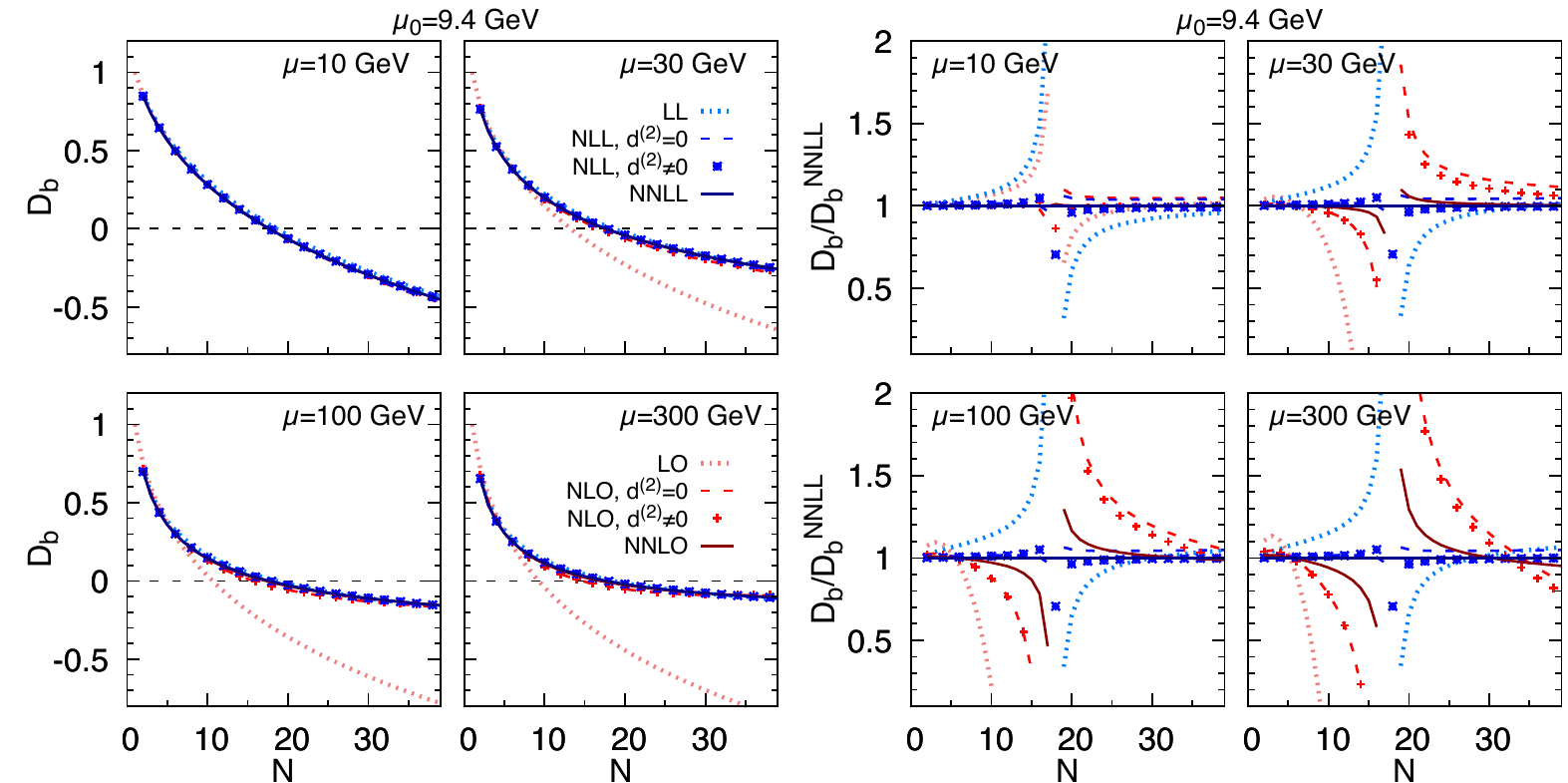}
\caption{\label{fig:DBtilde} The $N$-space bottom-quark fragmentation 
function, computed at different orders for the truncated solutions (LO, NLO, NNLO) and
resummed solutions (LL, NLL, NNLL). NLO and NLL solutions are shown both including (symbols)
and discarding (lines) the ${\cal O}(\as^2)$ terms in the initial conditions. 
Four values of the factorisation scale $\mu$ are considered in each panel: $\mu=10,\,30,\,100,\,300\,$
GeV. In the top (bottom) panels, the initial scale is set to
$\mu_0=m=4.7$ GeV ($\mu_0=9.4$ GeV). The right panels show the ratio 
over the NNLL-accurate predictions. }
\end{figure}

\begin{figure}[h!]
    \centering
    \includegraphics[width=\textwidth]{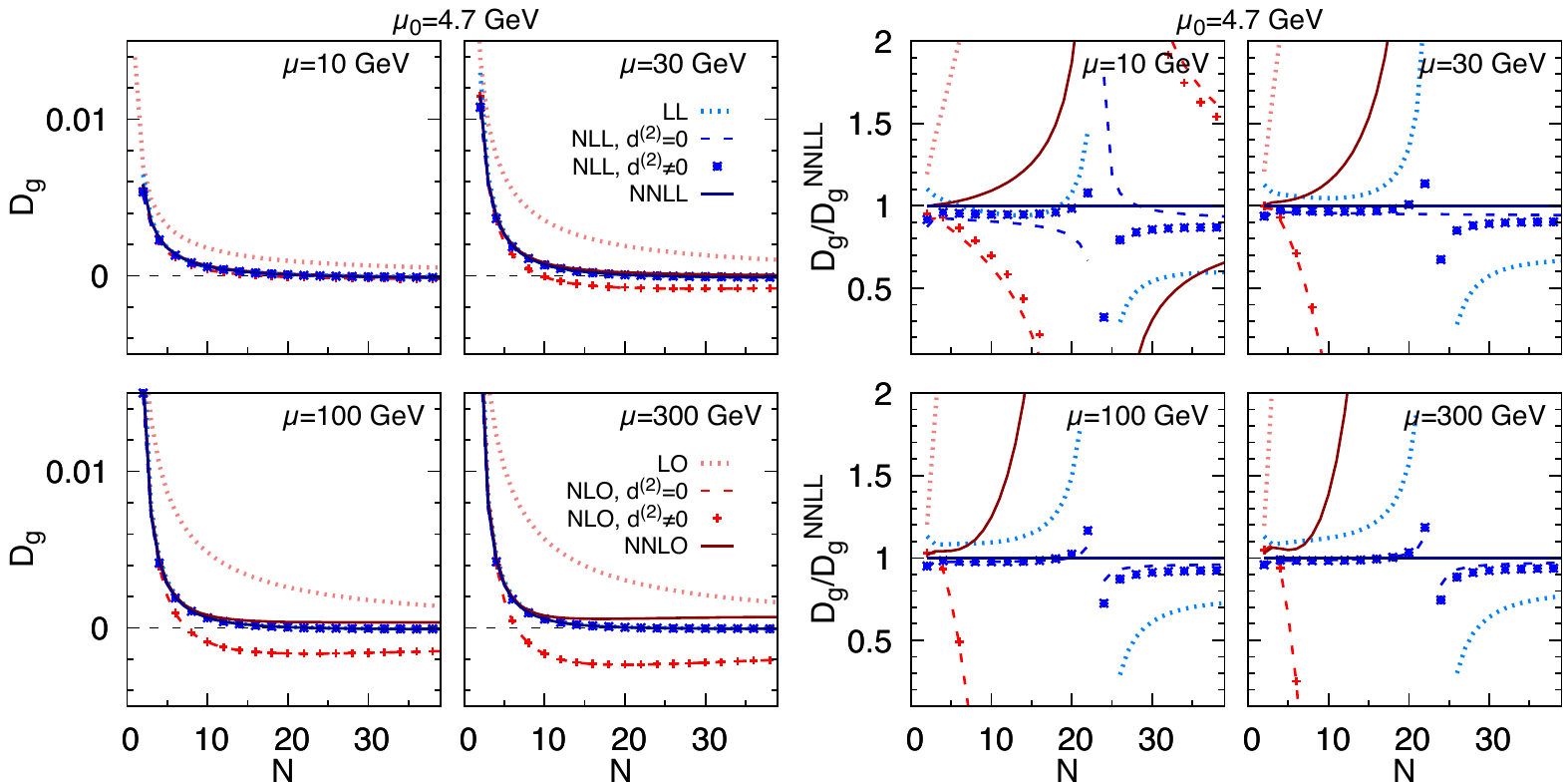}\\[5mm]
    \includegraphics[width=\textwidth]{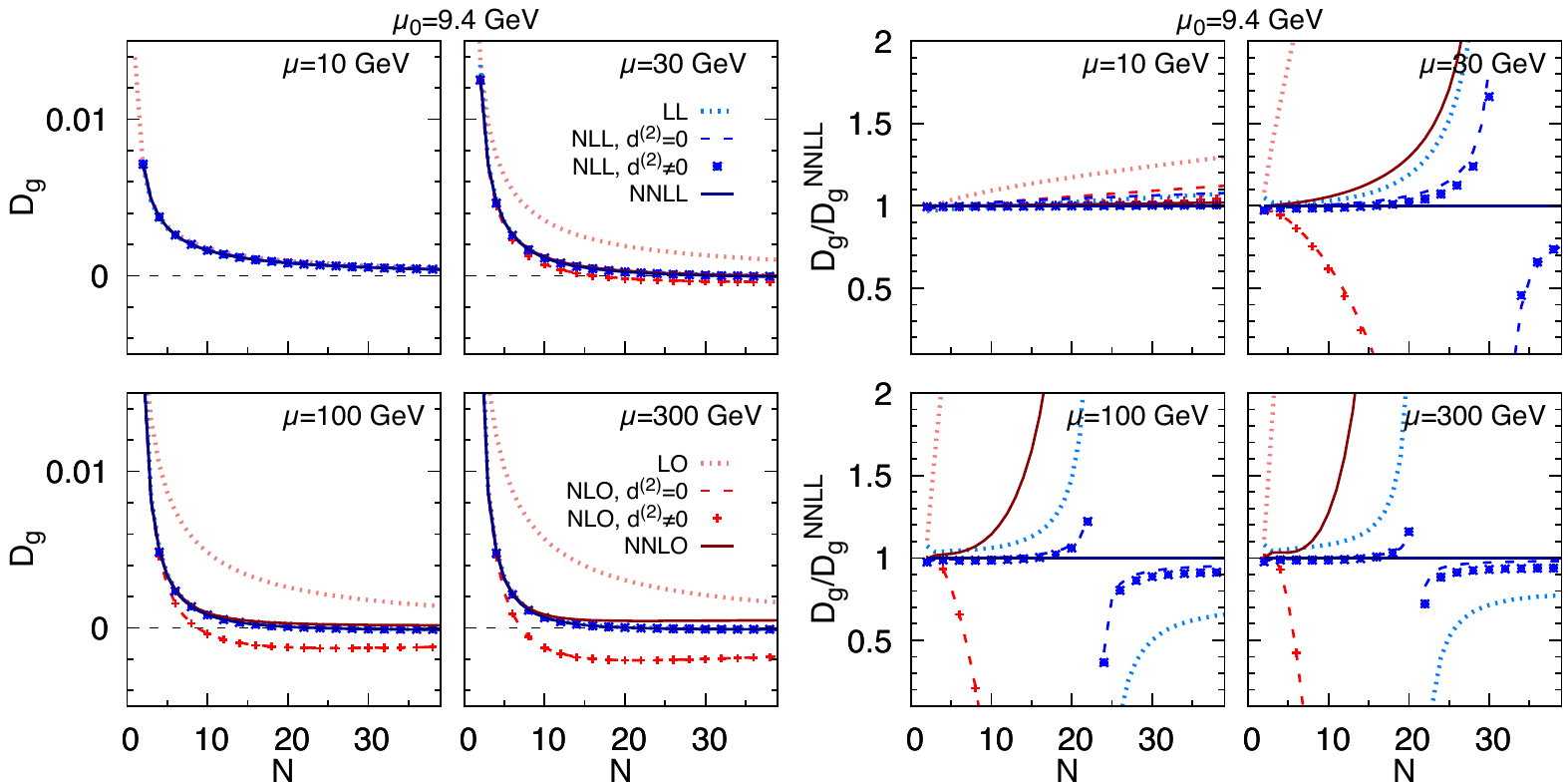}
\caption{\label{fig:DGtilde} Same as Fig.~\protect\ref{fig:DBtilde}, for the
  gluon fragmentation function.}
\end{figure}

We start by presenting results in $N$ space, for the $D_b$ and $D_g$ fragmentation functions, in Figs.~\ref{fig:DBtilde} and~\ref{fig:DGtilde} 
respectively. The layout of the figures is the following. Shades of red (blue), from lighter and more finely dashed to darker and solid, are used for truncated (resummed) predictions
of increasing perturbative orders, computed without initial conditions at order $\as^2$. Symbols are used for the NLO (NLL) predictions which
include the full initial conditions up to order $\as^2$. Left panels show results for the FFs, while right panels show the corresponding ratios w.r.t. the NNLL prediction.
In the top panels, an initial scale $\mu_0=m=4.7$ GeV is employed, while in the bottom ones it is set to $\mu_0=2m$. Finally, each panel shows results
at four different value of the scale $\mu$: $\mu=10,30,100,300$ GeV, from left to right and from top to bottom.

First, we inspect the behaviour of $D_b(N)$, displayed in Fig.~\ref{fig:DBtilde}: in the left plots, we can see how the resummed predictions
are hard to distinguish one from another, and also how the NNLO curve is close to them. This is not the case for the LO and NLO predictions: the LO
FF is clearly not apt at describing the bottom-quark initiated fragmentation, as it visibly departs from the resummed predictions already
for small or moderate values of $N$ ($N < 10$), and $\frac{\mu}{\mu_0}\simeq 2$. The NLO-accurate predictions show better agreement with the resummed ones. 

The $N$ range in Fig.~\ref{fig:DBtilde} is chosen to be $0\le N \le 40$ for illustrative purposes. However, this is not really consistent, because at very large values of $N$
(which correspond to values of $x$ close to 1) the initial conditions,
computed at a fixed order in $\as$, get large corrections from higher order contributions
due to the presence of large powers of $\log N$ in the perturbative coefficients.
The effect of such large logarithms is that the Mellin transforms of the
fragmentation functions in Fig.~\ref{fig:DBtilde} become negative around $N\sim 20$,
and consequently the ratio plots display a peak in that region.
This problem was pointed out in~\cite{Cacciari:2001cw}, where it is also argued
that a resummation of large $N$ logarithms in the initial condition would push the zero of the 
fragmentation functions toward much larger values of $N$. Even large-$N$ resummation, however, would not make the fragmentation functions positive in the whole range, 
due to non-perturbative
effects, or equivalently due to the presence of the Landau pole in the strong
coupling at very small energy scales. Perturbation theory, even in its resummed version, cannot provide a reliable description of fragmentation functions for $x$
larger than approximately $1-\Lambda_{QCD}/\mu$.

In the ratio plots the features described above can be appreciated with more details. In particular,
it can be appreciated how the NNLO-accurate prediction starts to depart from the NNLL at large scales and large values of $N$, while in general, with the
exception of the aforementioned spike, resummed computations show a better agreement with each other at large $N$. Increasing the
initial scale $\mu_0$, reduces the differences between the resummed and 
truncated predictions, as it is expected and as it was studied
in details in~\cite{Bertone:2017djs}. 
However the global pattern is unchanged. 
Finally, we observe how the impact of initial condition 
at order $\as^2$ is rather mild (at or below the 10\% level), 
regardless of the scale.

Turning to $D_g(N)$, in Fig.~\ref{fig:DGtilde}, we observe 
that the truncated expressions depart very quickly from the 
resummed ones, and how the
perturbative series wildly oscillates between negative and positive values for $N$ already as large as 8, with large differences from the resummed curves.
On the other hand, resummed curves lie rather close to each other (again, with the exception of the spike at $N=20$, induced by the initial conditions
of the quark FF), with differences that decrease with the scale, because of the running of $\as$. Differences are reduced when the initial scale
is doubled, and the impact of initial conditions is mild and rather constant with $N$.

\begin{figure}[h!]
    \centering
    \includegraphics[width=\textwidth]{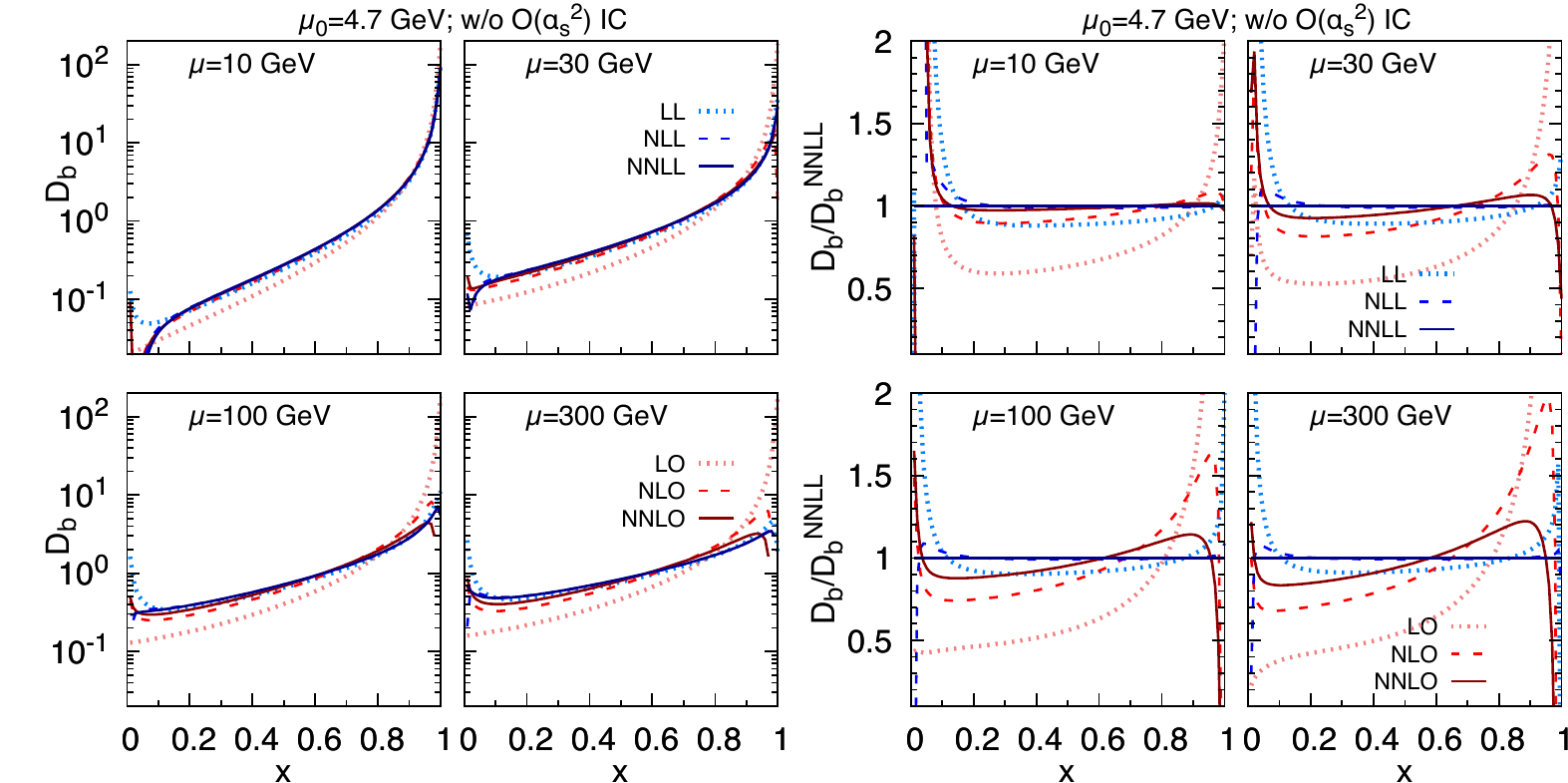}\\[5mm]
    \includegraphics[width=\textwidth]{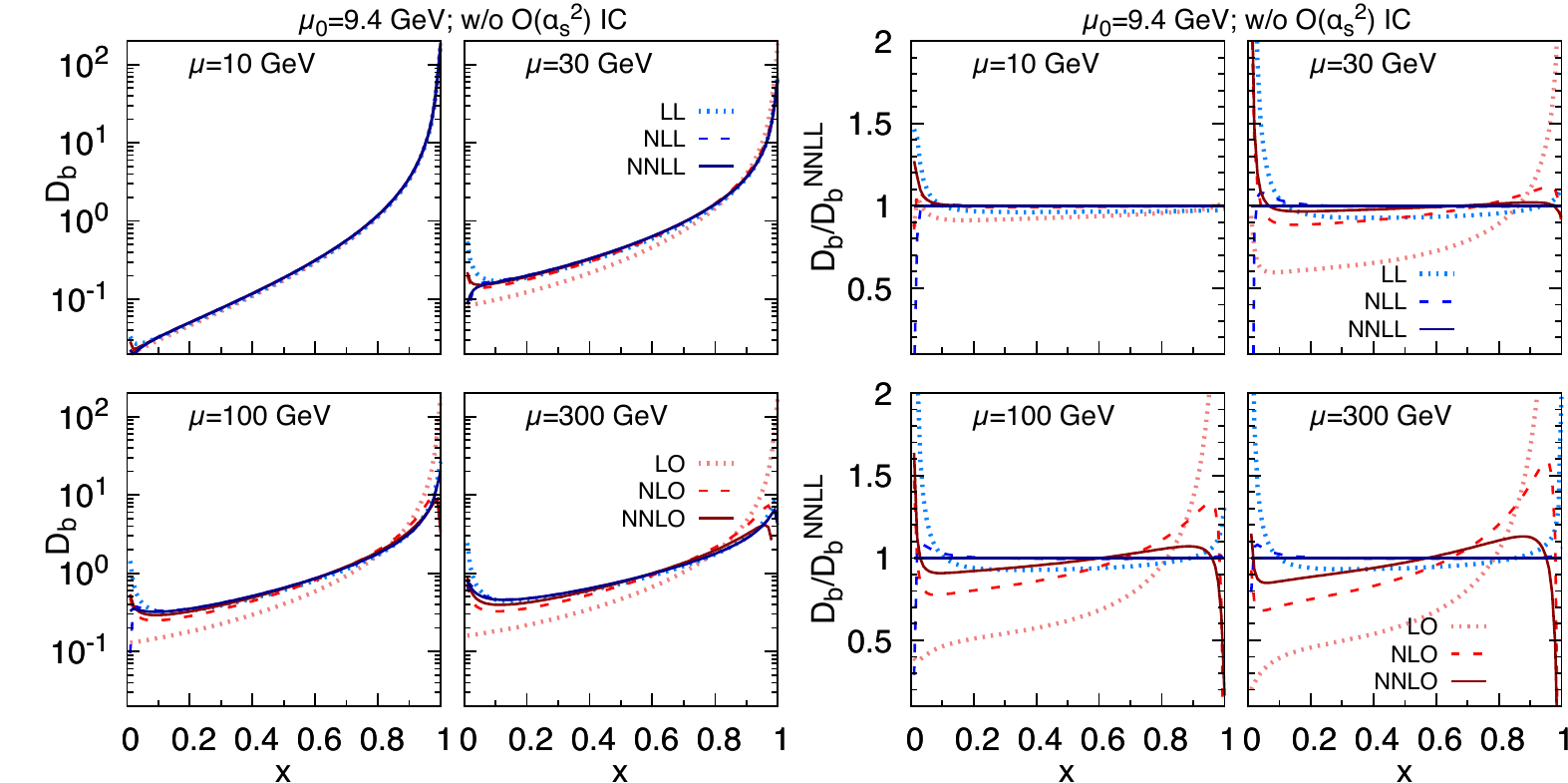}
\caption{\label{fig:DBXtilde} The $x$-space bottom-quark fragmentation 
function, computed at different orders for the truncated solution (LO, NLO, NNLO) and
resummed solutions (LL, NLL, NNLL). Four values of the factorisation
scale $\mu$ are considered in each panel: $\mu=10,\,30,\,100,\,300\,$
GeV. In the top (bottom) panels, the initial scale is set to
$\mu_0=m=4.7$ GeV ($\mu_0=9.4$ GeV). The right panels show the ratio 
over the NNLL-accurate predictions. }
\end{figure}

\begin{figure}[h!]
    \centering
    \includegraphics[width=\textwidth]{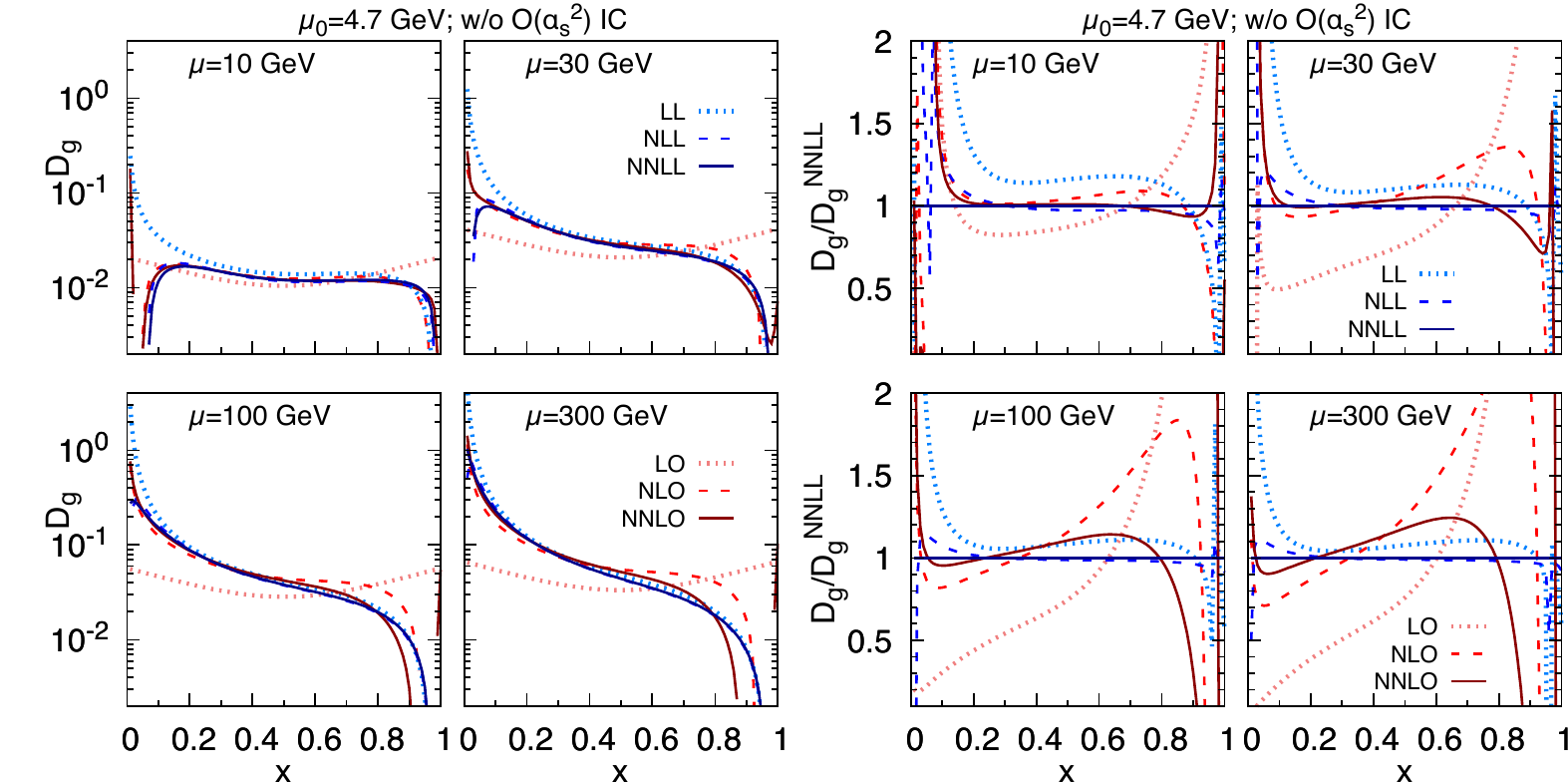}\\[5mm]
    \includegraphics[width=\textwidth]{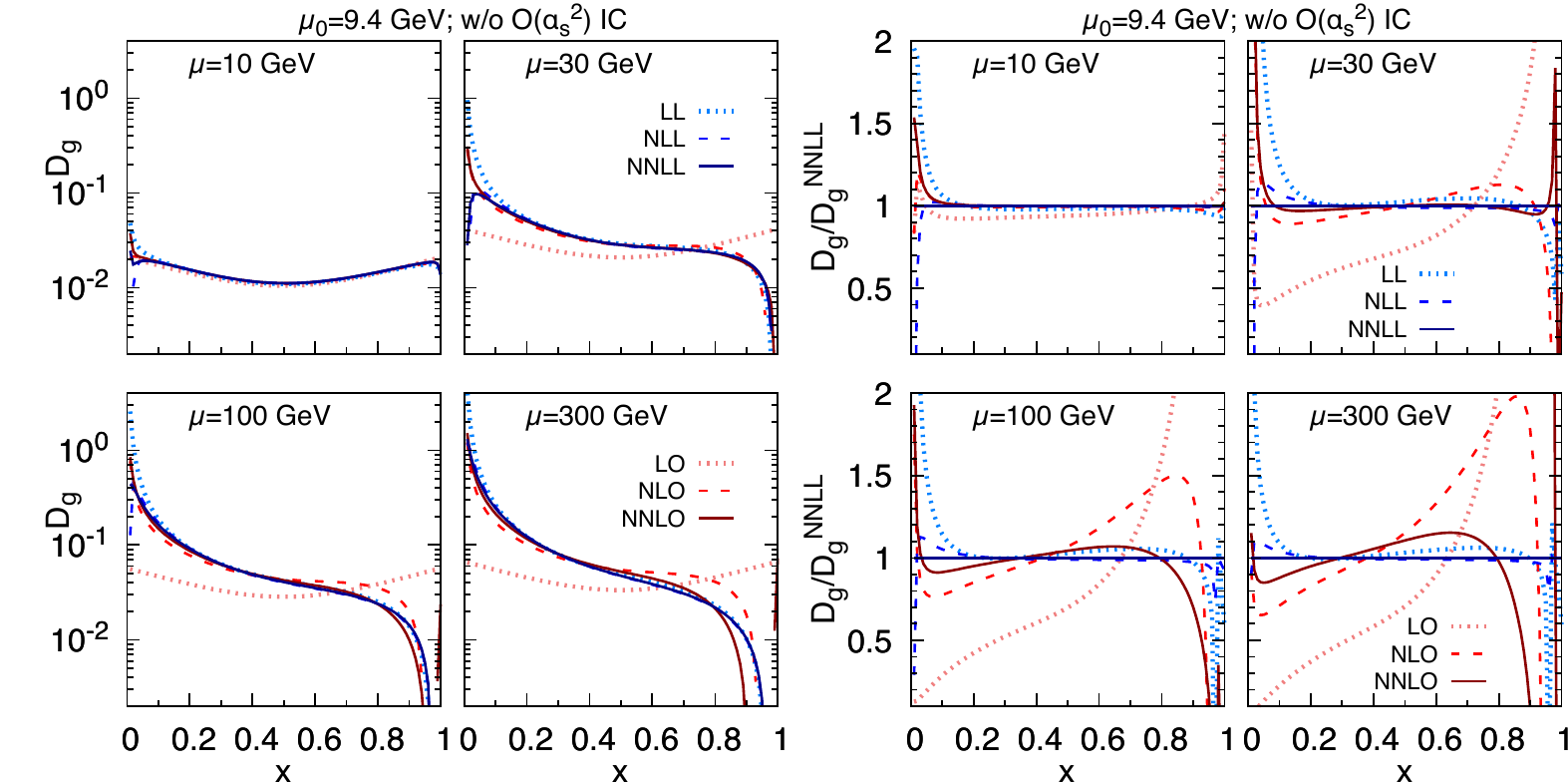}
\caption{\label{fig:DGXtilde} Same as Fig.~\protect\ref{fig:DBXtilde}, for the gluon fragmentation function.}
\end{figure}

In Figs.~\ref{fig:DBXtilde} and~\ref{fig:DGXtilde} the $x$-space results are displayed.
They generally reflect the pattern of the Mellin-space results, giving 
 a more direct feeling of the physics of the final-state splittings. Looking at $D_b(x)$,
in Fig.~\ref{fig:DBXtilde}, we appreciate how close the three resummed 
predictions are, for the four values of $\mu$ considered. Differences 
among the LL, NLL, and NNLL predictions are always within 10\% and with 
flat ratios, with the exception of very small and very
large $x$ ($x<0.1$ and $x>0.9$). The first
regime may only partially be accessible, since the physical regime is 
typically $x>\frac{m}{\mu}$. The behaviour in the second (large-$x$) regime
may be improved by resumming large-$x$ logarithms on top of the DGLAP ones~\cite{Cacciari:2001cw}. 
As far as the truncated predictions are concerned, 
they generally show a harder shape 
than the resummed ones (more steeply
peaked towards $x=1$), and the hardness decreases as higher orders are included. This is consistent with the fact that
higher order effects (i.e. extra radiations) soften the $b$ quark during the fragmentation, and in the case of resummed predictions these 
effects are included to all orders. If we take $\mu=100$ GeV as a representative scale, $\mu_0=4.7$ GeV, and consider the range $0.1<x<0.9$, the NLO-truncated 
prediction undershoots the NNLL resummed one of -25\% at small $x$, and overshoots it of +50\% at large $x$. At NNLO, differences are much reduced, at the level
of -10\% and + 15\%.

The gluon-initiated FF, $D_g(x)$, on the other hand, exhibits much larger 
differences between the truncated and resummed predictions. The most visible feature
is that the LO FFs is symmetric around $x=0.5$, while all the others are not. This is directly related to the 
symmetry of the $P_{qg}$ splitting function, which is
the only term at LO, as it can be seen from the first line of Eq.~\eqref{eq:truncG}\footnote{The initial condition $d_g^{(1)}$ is also proportional
to $P_{qg}$.}. 
As a consequence, the shape of the LO-truncated prediction does not change 
with the scale. Again, higher-order predictions
soften the shape of the splitting function, with rather dramatic effects both going from LO to NLO and from NLO to NNLO. 
In Sec.~\ref{sec:simul} we will show that these effects are dominantly due to the radiation from the parent gluon. 
As we did for $D_b(x)$,
considering the case $\mu=100$ GeV, $\mu_0=4.7$ GeV, it is apparent 
how the (N)NLO prediction exceeds the NNLL baseline by 80\% (15\%) at 
large $x$, and undershoots
it by -20\% (-5\%) at small $x$. Comparing resummed predictions among 
themselves shows, again, that the effect of sub-leading
logarithms is rather mild and, as anticipated by studying the 
behaviour in Mellin space, it is reduced when the scale $\mu$ is increased. 
Finally, some pathologic behaviour is visible both at small and large $x$. 
The latter
can likely be cured by resumming large-$x$ logarithms in the quark 
initial conditions~\cite{Cacciari:2001cw}, while for the former
small-$x$ resummation and coherence effects need to be considered, two ingredient which are crucial in order to obtain
correct predictions for heavy-quark multiplicities~\cite{Mangano:1992qq}, as we will discuss in Sec. \ref{sec:hqm}.
\begin{figure}[t!]
    \centering
    \includegraphics[width=\textwidth]{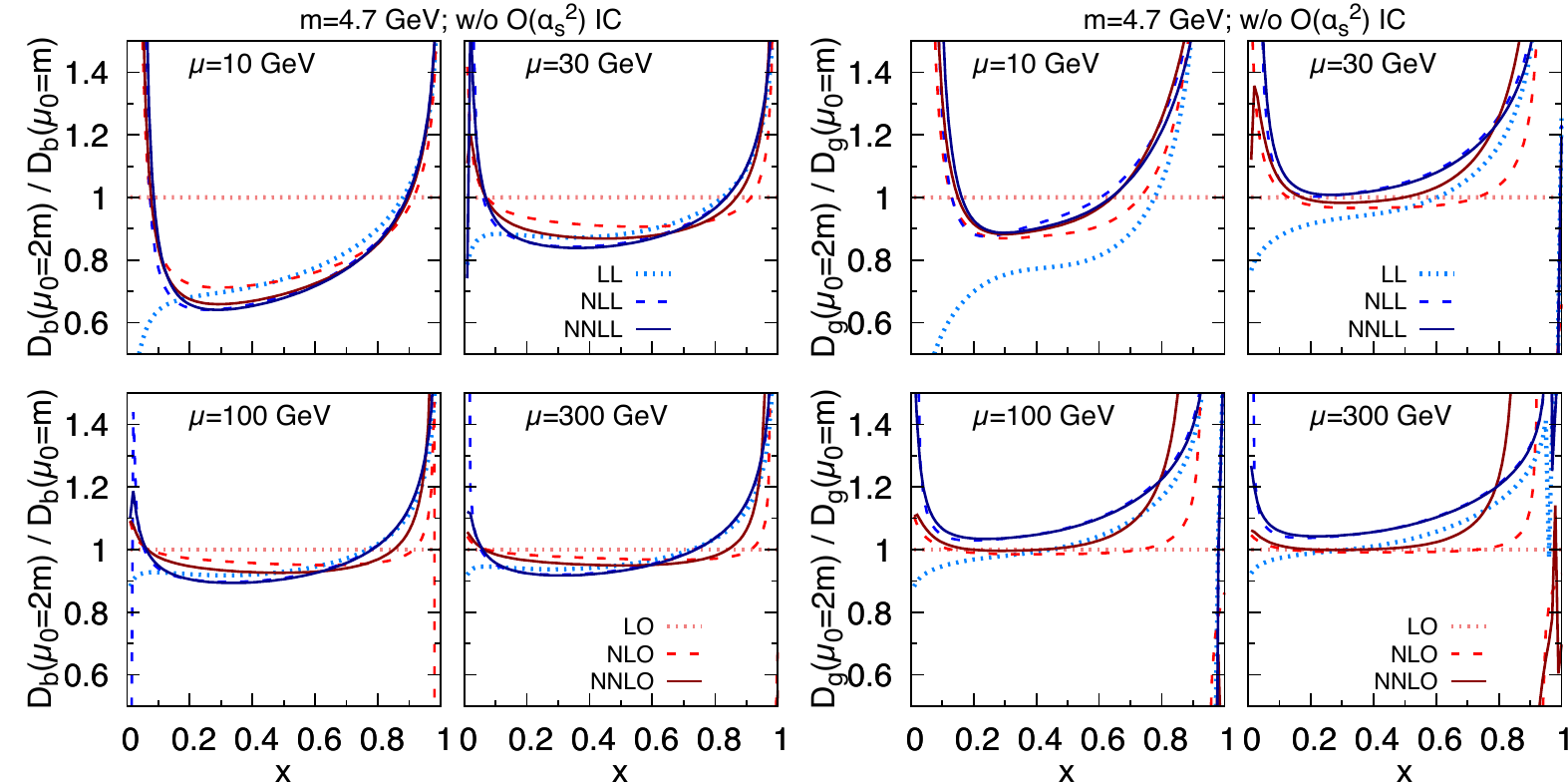}
\caption{\label{fig:DBGXtilde-mu0} The ratio $D_p(\mu_0=2m)/D_p(\mu_0=m)$ in $x$ space, with $p=b,g$ in the left and right 
panels respectively, computed at different orders for the truncated solution (LO, NLO, NNLO) and
resummed solutions (LL, NLL, NNLL). Four values of the factorisation
scale $\mu$ are considered in each panel: $\mu=10,\,30,\,100,\,300\,$
GeV.} 
\end{figure}

We conclude this section by discussing the dependence of truncated and resummed predictions 
on the initial scale $\mu_0$. The effect of changing the initial scale in the case of perturbatively generated bottom PDFs has been studied in details in \cite{Bertone:2017djs}. It is interesting to compare the effects in the case of 
perturbative bottom fragmentation functions. 
This is shown in Fig.~\ref{fig:DBGXtilde-mu0}, in $x$ space only, both for $D_b$ (left panels) and $D_g$ (right panels). 
In this figure we plot, for
each of the truncated and resummed predictions, the ratio $D_p(\mu_0=2m)/D_p(\mu_0=m)$, for the same values of $\mu$ as before. First,
we observe that at LO, no $\mu_0$ dependence is there, neither for $D_b$ nor for $D_g$. This can be easily understood by looking
at the initial conditions in Eqs.~\ref{eq:initb} and~\ref{eq:initg}, and at the 
truncated expressions in Eqs.~\ref{eq:truncB} and~\ref{eq:truncG}: the coefficient of the anomalous dimension, in both cases, will be
$\log{\frac{\mu_0^2}{m^2}}+\log{\frac{\mu^2}{\mu_0^2}}=\log{\frac{\mu^2}{m^2}}$. For the other predictions, both truncated and resummed, we
observe how the $\mu_0$ dependence is rather mild (less than 10\% for $D_b$ and 20\% for $D_g$ for $\mu\ge 100$ GeV) for intermediate 
values of $x$ and it decreases with the scale, because of the DGLAP evolution. Truncated predictions exhibit a more unstable behaviour at
large $x$, with a divergent structure in the pathologic region where $D_p(\mu_0=m)$ vanishes, and displaying 
larger uncertainties for higher perturbative orders. The same behaviour,
albeit with reduced $\mu_0$ dependence, is exhibited by resummed predictions. 
Overall, the $\mu_0$ dependence cannot be advocated
to explain the large differences between fixed-order and resummed predictions discussed earlier in this section.

\section{On the simulation of processes with $b$ quarks originated by timelike splittings}
\label{sec:simul}
\begin{figure}[h!]
    \centering
    \begin{overpic}[width=0.35\textwidth]{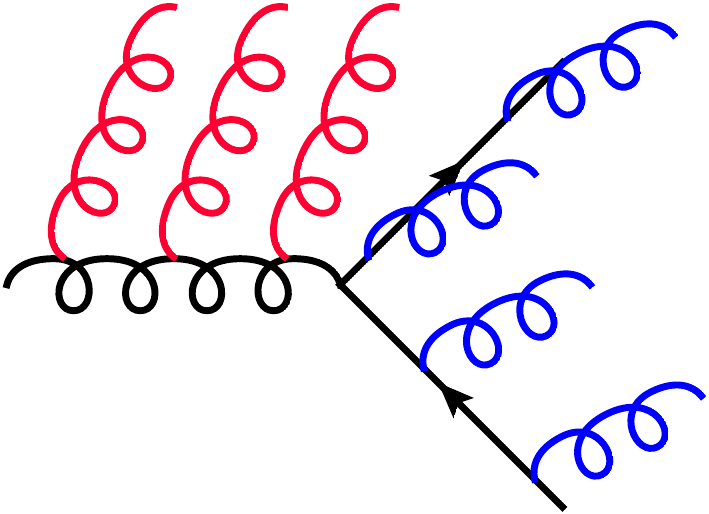}
        \put(10,10){\color{red} \huge $\sim P_{gg}$}
        \put(90,50){\color{blue} \huge $\sim P_{qq}$}
    \end{overpic}
\caption{\label{fig:gbb} The $g\to b\bar b$ splitting, dressed with extra gluon radiation. In the collinear limit, radiation off
the parent gluon (red) corresponds to factors of $P_{gg}$, while radiation off the quarks (blue) to factors $P_{qq}$.}
\end{figure}
\begin{figure}[h!]
    \centering
    \includegraphics[width=0.45\textwidth]{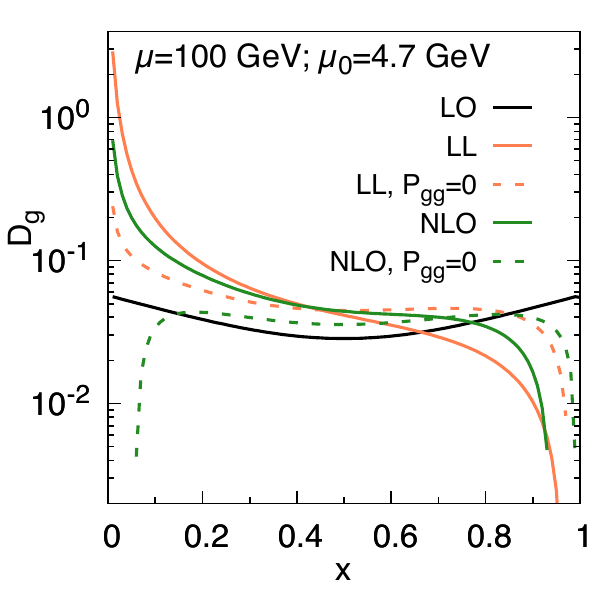}
\caption{\label{fig:Dgnopgg} The $D_g$ fragmentation function at LO (black), LL (orange) and NLO (green), for $\mu=100$ GeV, $\mu_0=4.7$ GeV. Solid
curves represent the complete expressions; dashed curves are obtained by setting $P_{gg}=0$.}
\end{figure}

\noindent
The results presented in Sec.~\ref{sec:results}, in particular those regarding the gluon-initiated FF, can provide instructive information
on the dynamics of final-state $g\to b\bar b$ splittings. As mentioned
in the introduction, such a mechanism is relevant for processes such as $b\bar b W$, 
 $y_t$-induced $b\bar b H$, $t\bar t b\bar b$ and multi-$b$ production. We can schematically represent the $g\to b\bar b$ splitting, including
 extra gluon emissions, as in Fig.~\ref{fig:gbb}. In that figure, the
 radiation off the parent gluon is shown in red, 
while the radiation off the  originating bottom quark is shown in
blue. 
Given the large effects observed in Sec.~\ref{sec:results}, a natural question to ask 
 is  whether the former or the latter type of radiation play a dominant role. At least
 two arguments can be used to show that the largest effects originate
 from the radiation off the gluon. The first argument 
 is related to color factors: in the collinear limit, each
 splitting from the parent gluon corresponds to a factor $P_{gg}$, 
proportional to $C_A$. Conversely, radiation off the quark corresponds to
 $P_{qq}$, proportional to $C_F$; since $C_A\simeq2C_F$, one expects
 the former effect to dominate over the latter. 
The second argument is that, 
 as it is visible in Fig.~\ref{fig:DBXtilde}, higher order effects distort the LO
 gluon-initiated FF 
towards small $x$, and $P_{gg}$ is the only splitting
 function which is singular in that regime. We support these arguments 
by explicitly showing, in Fig.~\ref{fig:Dgnopgg}, the NLO and LL predictions for
 $D_g(x)$ when setting $P_{gg}=0$, and comparing them to the full predictions (note that at NLO 
 -- second, third and fourth line of Eq.~\eqref{eq:truncG} -- the
 logarithmically-dominant term has 
either a single emission from the parent gluon, or
 one from the bottom quark). We choose $\mu=100$ GeV, $\mu_0=4.7$ GeV as a 
 representative example. We can clearly infer the importance 
of the emissions from the parent gluon, particularly
in the case of the NLO prediction. In that case, the single emission 
from the quark only mildly affects the symmetry of the FF. Also in the LL-resummed
 case, the prediction with $P_{gg}=0$ is much closer to the LO than to the complete LL prediction. 

These findings bear quite important consequences for the simulation of exclusive observables or in general observables
 sensitive to the $b$-quark degrees of freedom, in particular when predictions matched to parton shower are considered. 
 Since a parton shower radiates from external partons, if the bottom quarks
 appear in the hard-scattering process, only the radiation 
off the bottom quarks 
 (blue in Fig.~\ref{fig:gbb}) will be generated, while the radiation
 off the parent gluon will be included only at a given order 
in perturbation theory,
typically  NLO (with the exception of the results in~\cite{Buccioni:2019plc} which may be considered partly NNLO.) 
This is clearly not optimal. In general, resummed predictions exhibit a better 
 perturbative convergence with respect to finite-order calculations
 al ready at leading log. Hence,
in regimes dominated by the splitting mechanisms, it may be
 more appropriate to generate $b$ quarks by shower-evolving 
light partons, thus generating both kinds of radiation shown in Fig.~\ref{fig:gbb},
rather than to include them at the matrix-element level.~\footnote{
     The assumption that the shower is equivalent to our LL
     description holds up to small-$x$ effects. However, the largest differences
 between NLO and LL are at medium/large $x$.}
 
 Of course, some caveats must be considered. The above statement holds
 in case of exclusive observables, for which larger effects are expected. 
More inclusive observables (typically those related to the $b$-jet degrees of freedom)
 will display smaller effects. In Sec.~\ref{sec:hqm}, for example, we will
 show that this indeed the case for heavy quark multiplicities: the 
 effect of the resummation of final state collinear logarithms is much milder. 
Furthermore, an important assumption we are making is that 
fixed-order computations with a timelike $g\to b\bar b$ splitting
 follow the pattern of the FF at the corresponding order. 
This is certainly reasonable to assume, at least in those kinematic regions
in which the $g\to b\bar b$ splitting topology dominates. However
other effects must be taken into account: for example, 
mass effects typically affect the endpoint ($x\to1$ and $x\to0$)
behaviours of the FF, although not in a dramatic way. 
A further aspect is that the collinear approximation underlying a FF-based approach
 neglects the fact that the radiation recoil is spread
 among the other particles in the final state 
(see the extensive discussion
 in the case of $t\bar t b\bar b$ in~\cite{Buccioni:2019plc}). 
An assessment of the impact of these effects on physical observables requires a convolution
of fragmentation functions with a suitable partonic cross
section. This task is left for future work. 

\section{Relation with heavy quark multiplicities in gluon jets}
\label{sec:hqm}
Theoretical predictions for jet multiplicities, and in particular of heavy quark multiplicities, has witnessed an
important effort of the theory community during the 80's
(see e.g. \cite{Bassetto:1984ik,Mangano:1992qq} and references
therein). It is instructive to illustrate 
if, and in case how, FFs can be employed to predict such
multiplicities. 
We start by considering the main result of Ref.~\cite{Mangano:1992qq}, namely
the probability of a gluon with virtuality $Q^2$ to split into
a pair of quark-antiquark with mass $m$:
\begin{equation}
    \rho(Q^2) = \frac{1}{6 \pi} \int_{4 m^2} ^{Q^2} \frac{d K^2}{K^2} \as(K) \, n_g(Q^2,K^2) \left(1+\frac{2m^2}{K^2}\right) \sqrt{1-\frac{4m^2}{K^2}},
    \label{eq:hqm}
\end{equation}
where the gluon multiplicity $n_g$ is given by\footnote{With respect
  to the original Eq.~$1.2$ in~\cite{Mangano:1992qq},
we have replaced $\log\frac{Q^2}{\Lambda^2}$ by $\frac{1}{b_0\as(Q)}$ (and similarly for $\log\frac{K^2}{\Lambda^2}$).}
\begin{equation}
    n_g(Q^2,K^2)=\left(\frac{\as(K)}{\as(Q)}\right)^a
    \cosh\left[\sqrt{\frac{2 C_A}{\pi b_0}} \left(\frac{1}{\sqrt{b_0 \as(Q)}} - \frac{1}{\sqrt{b_0 \as(K)}} \right) \right].
    \label{eq:hqm-ng}
\end{equation}
The exponent $a$ has the value $a=-\frac{1}{4}\left[1+ \frac{2 C_A}{3\pi b_0} \left(1 - \frac{C_F}{C_A}\right)\right]$,
but it
can be set to zero as long as one's interest is restricted to the leading-logarithmic behaviour, which is the case
we are considering in this Section. The authors
of~\cite{Mangano:1992qq} mention that, in order to get the correct multiplicity, coherence effects
have to be accounted for in a systematic way. We will discuss the
corresponding effects in the case of FFs. 
We proceed by expanding Eq.~\eqref{eq:hqm} in powers of $\as(Q^2)$
neglecting all $m$ effects in the integrand, 
and by comparing such an expansion with the leading-logarithmic terms
in our truncation for $D_g$, Eq.~(\ref{eq:truncG}).
The expansion of Eq.~(\ref{eq:hqm}) to order $\as^2(Q)$ reads:
\begin{eqnarray}
    \rho(Q^2) &=& \frac{\as(Q)}{4\pi} \frac{2}{3} 
                        \log\frac{Q^2}{4 m^2}
                        \nonumber \\
        & +& \left(\frac{\as(Q)}{4\pi}\right)^2 \frac{2}{3} 
                        \left[2\pi b_0 \log^2 \frac{Q^2}{4 m^2}
                        + \frac{1}{3} C_A \log^3 \frac{Q^2}{4 m^2} \right].
    \label{eq:hqm-trunc}
\end{eqnarray}
This expression should be compared with the first moment
of the gluon fragmentation function, as given in  Eq.~(\ref{eq:truncG}),
expanded to second order in $\as(Q)$.
Keeping only the leading-log terms in Eq.~(\ref{eq:truncG})
we get
\begin{equation}
\lim_{N\to 1}\d_{g}(Q)= 
\lim_{N\to 1}\Bigg[\frac{\as(\mu)}{4\pi}
\frac{2}{3}t
+\left(\frac{\as(Q)}{4\pi}\right)^2\frac{2}{3}
\left( \frac{2C_A}{N-1}
+ 2 \pi b_0 \right) t^2\Bigg],
\label{eq:truncGLL1}
\end{equation}
where $t=\log\frac{Q^2}{\mu_0^2}$, and we have used  Eqs.~(\ref{eq:gammaqq}--\ref{eq:gammagg}) for $N=1$:
\begin{equation}
\gamma_{qq}^{(0)}(1)=0;\qquad
\gamma_{qg}^{(0)}(1)=\frac{4}{3}T_F=\frac{2}{3};
\qquad
\gamma_{gg}^{(0)}(N)=\frac{4C_A}{N-1}
+O\left((N-1)^0\right).
\end{equation}
We see that Eqs.~(\ref{eq:hqm-trunc}) and (\ref{eq:truncGLL1}) actually coincide,
with the choice $\mu_0=2m$,
apart from the term proportional to $C_A$, which is singular as $N\to 1$.
This singularity arises from the small-$x$ behaviour of the fragmentation function,
which diverges as $\frac{1}{x}$. One may regularise this singularity
by restricting the integration range to $x_{\rm min}\le x\le 1$,
with $x_{\rm min}$ of order $\frac{m^2}{Q^2}$.
This already provides the extra power of $\log\frac{m^2}{Q^2}$ which appears
in the $C_A$ term in Eq.~(\ref{eq:hqm-trunc}),
 but fails to reproduce the coefficient of the $C_A$ term in Eq.~(\ref{eq:hqm-trunc}). 
A refinement of this procedure is achieved by including a kinematical constraints  in the form
of a $x$ dependence of the scale argument
of the fragmentation function~\cite{Bassetto:1979nt,Amati:1980ch}. This is 
motivated by the observation that the virtuality of a particle scales approximately
as $x$ when it
decays into two bodies with energy fractions $x$ and $1-x$, in the $x\to 0$ limit. The corresponding integral reads
\begin{eqnarray}
    \int_{\frac{4m^2}{Q^2}} ^1 dx \,D_g(x,x Q^2) &\ni& 
        \left(\frac{\as(Q)}{4\pi}\right)^2 \frac{4}{3} C_A \int_{\frac{4m^2}{Q^2}} ^1 \frac{dx}{x} 
        \log^2\frac{x Q^2}{4 m^2} 
        \nonumber\\
     &=&  \left(\frac{\as(Q)}{4\pi}\right)^2 \frac{4}{9} C_A \log^3 \frac{Q^2}{4 m^2}, 
\end{eqnarray}
which is a factor 2 larger than the corresponding term in Eq.~(\ref{eq:hqm-trunc}). The origin of this residual discrepancy is due to
dynamic (rather than kinematic) effects related to color coherence and angular ordering, as it was
shown explicitly in~\cite{Mueller:1981ex}. In general, at order $\as^p(Q)$, an extra factor 
$2^{p-1}$ will appear in the most singular term ($\sim {C_A}^{p-1} \log^{2p-1}\frac{Q^2}{4 m^2}$) 
of the FF-based prediction for the heavy quark multiplicity~\footnote{
Following the procedure outlined above, we also verified that this is the case at NNLO.}. Color coherence effects
are not included in our framework, and require matching with small-$x$ resummation in order to be fully accounted for.

\begin{figure}[h!]
    \centering
    \includegraphics[width=0.95\textwidth]{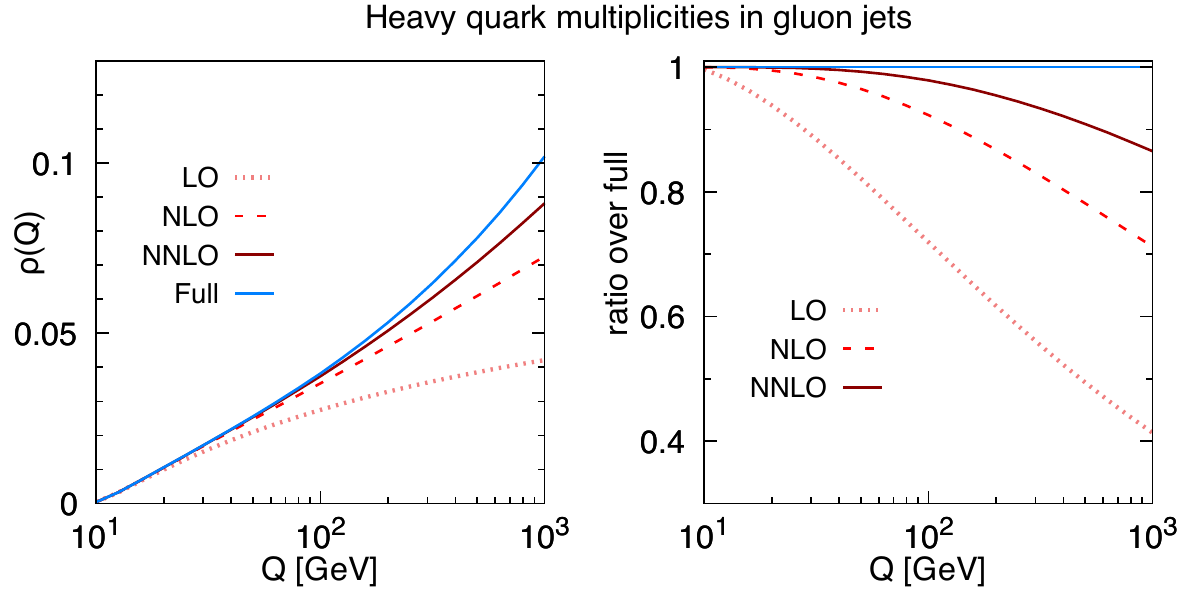}
    \caption{\label{fig:hqm} The heavy quark multiplicity computed by truncating Eq.~\protect\eqref{eq:hqm} up to different 
    orders of $\as(Q^2)$, compared with the full result. The right panel show the ratio over the full result.}
\end{figure}

We conclude this section by showing, in Fig.~\ref{fig:hqm}, the comparison of the truncation of Eq.~\eqref{eq:hqm} with the full result. The
effect of the truncation can be appreciated both by considering the absolute multiplicity (left panel), and the ratio 
 of the truncation up to order $\as(Q)$, $\as^2(Q)$ and $\as^3(Q)$ over the complete result. We do not set the exponent $a$ to zero in 
this case. Owing to the inclusiveness of this observable, effects are much milder than those
observed in Sec.~\ref{sec:results}: at $Q=100$~GeV, the LO prediction is about 70\% of the total, while
the NLO one approximates the total by less than 10\%\footnote{
 Some unpublished results in a presentation by S. Pozzorini confirm these numbers, see \\
 \url{https://indico.cern.ch/event/727396/contributions/3018606/attachments/1659440/2657963/benasque18.pdf}.
}. This indicates how the importance of the effects described in this paper changes
when inclusive observables are considered, and motivates further works
assessing the impact of resummation of collinear logarithm on
realistic cross sections.

\section{Conclusions and outlook}
\label{sec:conc}
One of the major obstacles to precision physics at the LHC and at future 
hadron colliders is currently given by our limited understanding of the
associate production of heavy quarks (typically bottom quarks) 
and heavier objects. Because of their multi-scale nature, the
description of such processes in perturbative QCD is highly
non-trivial. In particular, processes
where heavy quarks are dominantly produced via final-state splittings 
are affected by the largest theoretical uncertainties, both due to
missing higher orders and to parton-shower and matching systematics. 
In this paper, we make a step forward in the comprehension
of the dynamics underlying these processes. We adopt a FF-based
approach, which allows us to assess the impact of 
logarithms appearing at each order in perturbation theory, and
to establish the importance of their resummation. 
By considering truncated FFs up to NNLO 
in QCD, we mimic the description of a fixed-order computation for processes 
involving the corresponding splittings at the same order. We
investigated both bottom-initiated and gluon-initiated production of a
bottom quark. 
In both cases, and particularly for the latter, 
a fixed-order description at LO or NLO turns out not to be adequate, and
either NNLO effects must be included or collinear logarithms must be
resummed to all orders in order to get reliable predictions for the bottom quark kinematics. When
more inclusive observables are considered, these effects are (much) reduced, 
as it has been shown for the case of heavy-quark multiplicities.

While the limits of a fixed-order description have
been known for some time in the case of bottom-initiated splitting, 
which is relevant e.g. for heavy-flavour production at 
large transverse momenta, to the best of our knowledge 
this has never been investigated for the gluon-initiated heavy quark
production. We have discussed in details 
the implications of our findings in the choice of scheme to describe
this kind of processes.
Our analysis motivates the effort to develop techniques aimed at
combining calculations matched to parton shower, which retain
 the advantages of the 4F and the 5F schemes in the appropriate kinematics
 region, as it is currently being developed for example in~\cite{Hoche:2019ncc}. 
Furthermore, our study outlines both a similarity and a difference between the timelike and
the spacelike regimes. In particular, the evolution of a gluon from a 
high to a low scale (timelike) or from a low to a high one
(spacelike), is associated mostly with radiating gluons,
 and only eventually a gluon splits into a heavy-quark pair. This is
 true in both cases. The main
 difference is that, in the case of spacelike splitting, the 
heavy quark line enters the scattering process and therefore gluon emissions are
resummed by the evolution of parton distributions even in a 4FS. 
On the contrary, in the case of a timelike 
 splitting, the gluon is the particle linked with the rest of the scattering,
 hence no resummation of these emissions is performed.

To conclude, for processes dominated by final-state $g\to b\bar b$ splittings,
the dominant contribution comes from the radiation off the parent gluon, 
rather than off the bottom (anti\nobreakdash-) quark. 
A recent study on charm production corroborates these findings: in fact, in Ref.~\cite{Helenius:2018uul}
the authors comment on the importance of a proper 
treatment of final-state splittings and on issues related to NLO+PS simulations in a massive scheme (see Fig.~12 therein and the relative discussion).
As a result, simulations
where bottom quarks are treated at the matrix-element level might not
be the most adequate to accurately describe these processes, 
at least in phase-space regions in which these splittings are dominant. 
Despite a massive scheme (or more generally
a scheme where bottom quarks are generated at the hard matrix-element
level) is often advocated as superior with respect to a
massless one, thanks to the possibility it provides to describe the whole
phase space, without cuts,  we have shown that assuming
the smallness of collinear logarithms, analogously to what happens
with their initial-state counterpart, is not always correct and 
may yield serious flaws in the description of the kinematics of the $b$ quark.

This work has several natural follow-ups. First we will improve the description of
FFs by including the second-order initial-conditions in the $x$ space, and assess
the impact of large- and small-$x$ resummation. 
While these improvements will make our results more consistent, we do
not expect them to change the final picture in any dramatic way. 
Most importantly, we will assess the importance of final-state
collinear logarithms on a realistic process and compare a FF-based description (both
using resummed and truncated results) within a NLO-accurate computation
(the inclusion of FFs in NLO subtraction schemes can be found in Refs.~\cite{Catani:1996vz,Frederix:2018nkq})
with a description at fixed order in QCD and possibly with data. While typical
analyses for processes like $W b\bar b$ and $t\bar t b\bar b$ study $b$-jet 
observables (for the latter, see the recent analyses~\cite{Aaboud:2018eki,Sirunyan:2019jud}), it is not 
unreasonable to expect that more exclusive quantities related
to the bottom-flavoured hadrons will be measured, specially with the larger statistics
of the upcoming LHC runs.

\section*{Acknowledgements}
We are indebted with Fabio Maltoni for discussions, for a critical reading of our manuscript and for collaboration at earlier 
stages of this work. We thank Richard Fitton for the work done at the
initial stage during his Masters' thesis with M.~U. 
We would like to thank Stefano Pozzorini and Laura Reina for having suggested to consider 
heavy quark multiplicities in our study, and Paolo Nason for enlightening discussions on this 
matter. We are also grateful to Alex Mitov for discussion about the NNLO initial conditions and for providing us with
the corresponding machine-readable expressions. Discussions on Mellin transforms 
with Johannes Bluemlein, Paolo Torrielli, Jos Vermaseren, and Andreas Vogt are
gratefully acknowledged. We would like to specially thank Johannes Bluemlein for sharing 
the still-unpublished computer code {\tt ancont1}.
G.~R. is partially supported by a PRIN 2017 project of the Italian Ministry of University and Research.
GR is supported by the Italian Ministry of Research (MIUR) under grant PRIN 20172LNEEZ. 
M.~U. is partially supported by the STFC consolidated grant ST/L000681/1 and the
Royal Society grant RGF/EA/180148. She is funded by the Royal Society grant DH/150088. 
M.~Z. has been supported in part by the Netherlands National Organisation for Scientific Research (NWO).

\appendix

\section{Truncated expressions and validation}
\label{app:truncate}
In this Appendix we give explicit expressions for the truncation of the FFs up to oder $\as^3$ and discuss the method we used to validate numerically our analytic expressions.

\subsection{Non-singlet solution}
Expanding both Eq.~\eqref{eq:unnlo} and the initial conditions up to
${\cal O}(\as^3)$, and omitting the $N$ dependence everywhere and the $m$ 
dependence from the coefficients of the initial conditions, we obtain 
the following truncated solutions:
{\small
\begin{align}
\d_{T_{24}}(\mu) &= - (\nf -1) \Bigg\{
1+\frac{\as(\mu)}{4\pi}\left[d_{b}^{(1)}(\mu_0)+\gamma_+^{(0)}t\right]\notag\\
& +\left(\frac{\as(\mu)}{4\pi}\right)^2
\bigg[d_{b}^{(2)}(\mu_0) + d_{\bar{b}}^{(2)}(\mu_0)- 2 d_{q}^{(2)}(\mu_0) 
\notag\\
& \qquad + \left(\gamma_+^{(1)} + d_{b}^{(1)}(\mu_0)\gamma_+^{(0)} 
+4\pi b_0 d_{b}^{(1)}(\mu_0)\right) t 
+ \left(\frac{1}{2}(\gamma_+^{(0)})^2 + 2 \pi b_0 \gamma_+^{(0)}\right)t^2\bigg]
\notag\\
& +\left(\frac{\as(\mu)}{4\pi}\right)^3
\bigg[
\bigg(\left(d_{b}^{(2)}(\mu_0)+d_{\bar{b}}^{(2)}(\mu_0)-2d_{q}^{(2)}(\mu_0)\right)
\left(\gamma_{+}^{(0)}+8\pi b_0\right)\notag\\
&\qquad  +\gamma_{+}^{(2)} +16\pi^2b_1 d_{b}^{(1)}(\mu_0)
+d_{b}^{(1)}(\mu_0)\gamma_{+}^{(1)} \bigg)t 
\notag\\
& \qquad  +\left(4\pi b_0\gamma_{+}^{(1)} + d_{b}^{(1)}(\mu_0)
\left(\frac{1}{2}(\gamma_{+}^{(0)})^2+6 \pi b_0 \gamma_{+}^{(0)} + 16\pi^2b_0^2\right)
+\gamma_+^{(0)}(\gamma_{+}^{(1)}+8 \pi^2 b_1)\right)t^2
\notag\\
&  \qquad  +\left(\frac{1}{6}(\gamma_{+}^{(0)})^3
+2\pi b_0 (\gamma_+^{(0)})^2+\frac{16}{3}\pi^2 b_0^2\gamma_+^{(0)}\right)t^3\bigg]
\Bigg\},
\label{eq:truncT24}
\end{align} }%
and
{\small
\begin{align}
\d_{V_{b}}(\mu) &= 1+\frac{\as(\mu)}{4\pi}
\left[d_b^{(1)}(\mu_0)+\gamma_V^{(0)}t\right]
\notag\\
&+\left(\frac{\as(\mu)}{4\pi}\right)^2 \bigg[ d_b^{(2)}(\mu_0) - d_{\bar{b}}^{(2)}(\mu_0)\notag\\
&  \qquad
+\left(\gamma_V^{(1)}+d_b^{(1)}(\mu_0)\gamma_V^{(0)} +4\pi b_0 d_b^{(1)}(\mu_0)\right)t 
+ \left(\frac{1}{2}(\gamma_V^{(0)})^2 + 2\pi b_0  \gamma_V^{(0)}\right)t^2\bigg]
\notag\\
&+\left(\frac{\as(\mu)}{4\pi}\right)^3
\bigg[
\bigg(\left(d_{b}^{(2)}(\mu_0)-d_{\bar{b}}^{(2)}(\mu_0)\right)\left(\gamma_V^{(0)}
+8\pi b_0\right)
\notag\\
& \qquad  
+\gamma_V^{(2)} +16\pi^2 b_1d_{b}^{(1)}(\mu_0)+d_{b}^{(1)}(\mu_0)\gamma_V^{(1)} \bigg)t \notag\\
& \qquad  
+\left(4\pi b_0\gamma_V^{(1)} + d_{b}^{(1)}(\mu_0)\left(\frac{1}{2}(\gamma_V^{(0)})^2
 + 6 \pi b_0 \gamma_V^{(0)} + 16\pi^2 b_0^2 \right)+\gamma_V^{(0)}(\gamma_V^{(1)}
 +8\pi^2 b_1)\right)t^2
\notag\\
& \qquad  +\left(\frac{1}{6}(\gamma_V^{(0)})^3 
+ 2\pi b_0(\gamma_V^{(0)})^2 + \frac{16}{3}\pi^2 b_0^2\gamma_V^{(0)}\right)t^3\bigg].
\label{eq:truncVb}
\end{align} }%

\subsection{Singlet solution}
Expanding both Eq.~\eqref{eq:unnlo} and the initial conditions up to
${\cal O}(\as^3)$, and omitting the $N$ 
dependence everywhere and the $m$ 
dependence from the coefficients of the initial conditions, we obtain 
the following truncated solutions\footnote{Note that the products in Eq.~\eqref{eq:unnlo} are products of $2\times 2$ matrices.}
{\small
\begin{align}
\d_{\Sigma}(\mu) &=
1+\frac{\as(\mu)}{4\pi}\left[d_{b}^{(1)}(\mu_0)+\gamma_{qq}^{(0)}t\right]\notag\\
& +\left(\frac{\as(\mu)}{4\pi}\right)^2
\bigg[d_{b}^{(2)}(\mu_0) + d_{\bar{b}}^{(2)}(\mu_0) + 2(\nf -1) d_{q}^{(2)}(\mu_0) 
\notag\\
& \qquad +\left(d_{b}^{(1)}(\mu_0)\gamma_{qq}^{(0)}  +  d_{g}^{(1)}(\mu_0)\gamma_{gq}^{(0)} +\gamma_{qq}^{(1)} +4\pi b_0d_{b}^{(1)}(\mu_0) \right)t 
\notag\\
& \qquad + \left(\frac{1}{2}(\gamma_{qq}^{(0)})^2 
+ \frac{1}{2}\gamma_{qg}^{(0)}\gamma_{gq}^{(0)}+ 2 \pi b_0 \gamma_{qq}^{(0)}\right)t^2  \bigg] 
\notag\\
& +\left(\frac{\as(\mu)}{4\pi}\right)^3
\Bigg[
\Bigg(\bigg(d_b^{(2)}(\mu_0) + d_{\bar{b}}^{(2)}(\mu_0) + 2(\nf -1) d_q^{(2)}(\mu_0)\bigg)
\left(\gamma_{qq}^{(0)}+8\pi b_0\right) 
\notag\\
&  \qquad  
+ d_g^{(2)}(\mu_0)\gamma_{gq}^{(0)}
+\gamma_{qq}^{(2)}+16\pi^2 b_1 d_b^{(1)}(\mu_0)+d_b^{(1)}(\mu_0)\gamma_{qq}^{(1)}
+d_g^{(1)}(\mu_0)\gamma_{gq}^{(1)}\Bigg)t 
\notag\\
&  \qquad
+\Bigg(4\pi b_0\gamma_{qq}^{(1)} 
+ d_b^{(1)}(\mu_0)\left(\frac{1}{2}(\gamma_{qq}^{(0)})^2
+ 6 \pi b_0 \gamma_{qq}^{(0)} + 16\pi^2 b_0^2 +\frac{1}{2}\gamma_{gq}^{(0)} \gamma_{qg}^{(0)} \right) 
\notag\\
&  \qquad 
+\gamma_{qq}^{(0)}(\gamma_{qq}^{(1)}+8\pi^2b_1 )
+ \frac{1}{2} \gamma_{qg}^{(0)}\gamma_{gq}^{(1)}
+ \frac{1}{2}\gamma_{gq}^{(0)}\gamma_{qg}^{(1)} 
+ \frac{1}{2}d_g^{(1)}\gamma_{gq}^{(0)}\left(\gamma_{gg}^{(0)}+\gamma_{qq}^{(0)}
+12\pi b_0\right)\Bigg)t^2
\notag\\
&  
+\Bigg(\frac{1}{6}(\gamma_{qq}^{(0)})^3 
+ 2\pi b_0(\gamma_{qq}^{(0)})^2 + \frac{16}{3}b_0^2\pi^2\gamma_{qq}^{(0)}
+ \frac{1}{6} \gamma_{qg}^{(0)}\gamma_{gq}^{(0)}
\left(\gamma_{gg}^{(0)} + 2\gamma_{qq}^{(0)} +12\pi b_0\right)\Bigg)t^3
\Bigg],
\label{eq:truncSing}
\end{align} }%
and
{\small
\begin{align}
\d_{g}(\mu) &= \frac{\as(\mu)}{4\pi}\left[d_{g}^{(1)}(\mu_0)+\gamma_{qg}^{(0)}t\right] 
\nonumber\\
&+\left(\frac{\as(\mu)}{4\pi}\right)^2
\bigg[ d_{g}^{(2)}(\mu_0) 
+ \left( d_{b}^{(1)}(\mu_0)\gamma_{qg}^{(0)} + d_{g}^{(1)}(\mu_0)\left(\gamma_{gg}^{(0)}+ 4\pi b_0  \right) + \gamma_{qg}^{(1)} \right) t 
\notag\\
& \qquad + \left( \frac{1}{2}\gamma_{gg}^{(0)}\gamma_{qg}^{(0)}
+ \frac{1}{2}\gamma_{qg}^{(0)}\gamma_{qq}^{(0)}
+ 2 \pi b_0 \gamma_{qg}^{(0)}\right) t^2\bigg]
\notag\\
&+\left(\frac{\as(\mu)}{4\pi}\right)^3
\Bigg[
\Bigg(\bigg(d_b^{(2)}(\mu_0) + d_{\bar{b}}^{(2)}(\mu_0) + 2(\nf -1) d_q^{(2)}(\mu_0)\bigg)
\gamma_{qg}^{(0)} 
+8\pi b_0 d_g^{(2)}(\mu_0) + d_g^{(2)} \gamma_{gg}^{(0)} \notag\\
&  \qquad 
+\gamma_{qg}^{(2)}+16\pi^2b_1 d_g^{(1)}(\mu_0)+d_b^{(1)}(\mu_0)\gamma_{qg}^{(1)}+ d_g^{(1)}(\mu_0)\gamma_{gg}^{(1)}\Bigg)t 
\notag\\
&  \qquad 
+\Bigg(4\pi b_0\gamma_{qg}^{(1)} + d_b^{(1)}(\mu_0)\gamma_{qg}^{(0)}\left(\frac{1}{2}\gamma_{gg}^{(0)}
+\frac{1}{2}\gamma_{qq}^{(0)} +6\pi  b_0 \right) \notag\\
&  \qquad 
+ d_g^{(1)}(\mu_0)\left(\frac{1}{2}(\gamma_{gg}^{(0)})^2+\frac{1}{2}\gamma_{gq}^{(0)}\gamma_{qg}^{(0)}+6\pi b_0\gamma_{gg}^{(0)}+16\pi^2 b_0^2\right)
\notag\\
&  \qquad 
+\gamma_{qg}^{(0)}\left(\frac{1}{2}\gamma_{gg}^{(1)}+\frac{1}{2}\gamma_{qq}^{(1)}
+8\pi^2 b_1 \right)
+\gamma_{qg}^{(1)}\left(\frac{1}{2}\gamma_{gg}^{(0)}+\frac{1}{2}\gamma_{qq}^{(0)}\right)\Bigg)t^2\notag\\
&  \qquad 
+\frac{1}{6} \gamma_{qg}^{(0)}\Bigg( (\gamma_{gg}^{(0)})^2
+\gamma_{gq}^{(0)}\gamma_{qg}^{(0)} + (\gamma_{qq}^{(0)})^2+
\gamma_{gg}^{(0)}\gamma_{qq}^{(0)} 
+ 12 \pi b_0 (\gamma_{qq}^{(0)}+\gamma_{gg}^{(0)}) + 32\pi^2 b_0^2 \Bigg)t^3
\Bigg].
\label{eq:truncG}
\end{align} }%

\subsection{Bottom quark}
Summing up the singlet and non-singlet combinations according to Eq.~\eqref{eq:combination}, we get
{\small
\begin{align}
\d_{b}(\mu) &= 1 + \frac{\as(\mu)}{4\pi}\left[d_b^{(1)}(\mu_0)+\gamma_{qq}^{(0)}t\right]
\nonumber\\
& +\left(\frac{\as(\mu)}{4\pi}\right)^2
\bigg[d_b^{(2)}(\mu_0)  
 + \left(d_b^{(1)}(\mu_0)\gamma_{qq}^{(0)} 
+ \frac{1}{2\nf}d_g^{(1)}(\mu_0)\gamma_{gq}^{(0)} + \bar\gamma_{qq}^{(1)} +4\pi b_0 d_b^{(1)}(\mu_0) \right) t 
\notag\\
& \qquad +\left( \frac{1}{2}(\gamma_{qq}^{(0)})^2 +
  \frac{1}{4\nf}\gamma_{qg}^{(0)}\gamma_{gq}^{(0)} + 2\pi b_0
  \gamma_{qq}^{(0)}\right)t^2\bigg]
  \notag\\
&+\left(\frac{\as(\mu)}{4\pi}\right)^3
\Bigg[
\Bigg(
\bar\gamma_{qq}^{(2)} + d_b^{(1)}(\mu_0)\bar\gamma_{qq}^{(1)}
+  d_b^{(2)}(\mu_0) (8 \pi  b_0+\gamma_{qq}^{(0)}) 
+\frac{1}{2\nf} d_g^{(2)}(\mu_0)  \gamma_{gq}^{(0)}
\notag\\
&  \qquad +16 \pi ^2 b_1d_b^{(1)}(\mu_0) 
+\frac{1}{2\nf}d_g^{(1)}(\mu_0) \gamma_{gq}^{(1)} \Bigg)t 
\notag\\
&  \qquad 
+\Bigg( \left(4\pi b_0+\gamma_{qq}^{(0)}\right) \bar\gamma_{qq}^{(1)}  
+ d_b^{(1)}(\mu_0)\left(\frac{1}{4\nf}\gamma_{gq}^{(0)}\gamma_{qg}^{(0)}
+\frac{1}{2}(\gamma_{qq}^{(0)})^2
 +6\pi b_0 \gamma_{qq}^{(0)}+16\pi^2b_0^2\right) 
 \notag\\
& \qquad 
+ \frac{1}{2\nf}d_g^{(1)}(\mu_0)\gamma_{gq}^{(0)}\left(\frac{1}{2}\gamma_{gg}^{(0)}
+\frac{1}{2}\gamma_{qq}^{(0)}+6\pi b_0\right)
+\frac{1}{4\nf}\gamma_{qg}^{(0)} \gamma_{gq}^{(1)}
+\frac{1}{4\nf}\gamma_{gq}^{(0)} \gamma_{qg}^{(1)} 
+ 8\pi^2 b_1 \gamma_{qq}^{(0)}\Bigg)t^2
\notag\\
& \qquad  
+
\frac{1}{6} \left(\gamma_{qq}^{(0)} \left(32 \pi ^2 b_0^2+12 \pi  b_0
   \gamma_{qq}^{(0)}+(\gamma_{qq}^{(0)})^2\right)
 +\frac{1}{2\nf}\gamma_{gq}^{(0)} \gamma_{qg}^{(0)} 
 \left(12 \pi  b_0+2\gamma_{qq}^{(0)}+\gamma_{gg}^{(0)}\right)
 \right)
t^3
\Bigg],
\label{eq:truncB}
\end{align} }%
where we have defined
\begin{equation}
\bar\gamma_{qq}^{(p)} = \frac{1}{2\nf} \left[\gamma_{qq}^{(p)} + (\nf -1)\gamma_+^{(p)} + \nf \gamma_V^{(p)}\right].
\end{equation}

\subsection{Validation}

\begin{figure}[ht]
    \centering
    \includegraphics[width=0.8\textwidth]{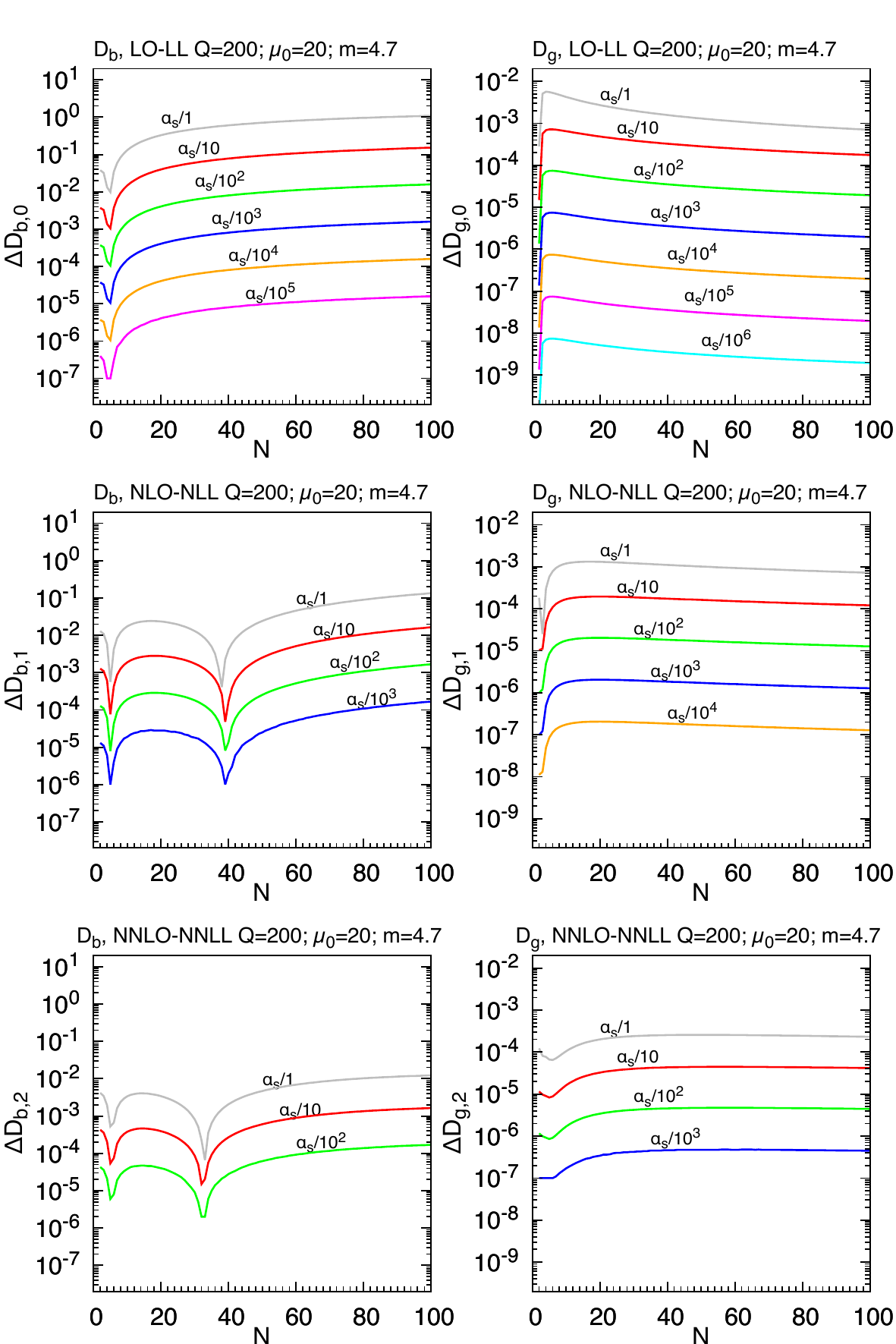}
    \caption{\label{fig:ascheck} The scaling of the difference $\Delta D_{p, q}$, defined in Eq.~\protect\eqref{eq:asdiff}, w.r.t. $\as$, 
    for $p=b,g$ (left, right) and $q=0,1,2$ (top, medium and bottom row).}
\end{figure}

To conclude this appendix, we discuss how the above expressions were validated in our computer code. 
We base our validation on two arguments: first, given an initial
condition, {\tt MELA} can provide  the evolution up to NNLL accuracy; second,
the difference 
\begin{equation}
    \Delta D_{p, q} \equiv \frac{\left|D_{p, {\rm N}^q{\rm LO}} -D_{p, {\rm N}^q{\rm LL}}\right|}{\as^{q+1}},
    \label{eq:asdiff}
\end{equation}
where $p=b,g,\ldots$ and $q=0,1,2$,
should be of $\mathcal O(\as)$. Thus, by changing the value of
$\as(m_Z)$, $\Delta D_{p, q}$ must display 
the same scaling. We show this scaling
in Fig.~\ref{fig:ascheck}, for $D_b$ (left) as well as $D_g$ (right), in $N$ space, for $q=0,1,2$ respectively in the top, central and bottom 
row. We fix the scales to $\mu=200$~GeV, $\mu_0=20$~GeV, 
and the bottom mass to $m=4.7$~GeV (in particular, by choosing $\mu_0\neq m$, all initial conditions are non-zero).

\providecommand{\href}[2]{#2}\begingroup\raggedright\endgroup


\end{document}